\documentclass[aps,prd,reprint,balancelastpage,floatfix,nofootinbib,preprintnumbers,amsmath,amssymb,superscriptaddress]{revtex4-1}

\usepackage[
    colorlinks=true,
    citecolor=blue,
    linkcolor=blue,
    urlcolor=blue
]{hyperref}

\usepackage{bm}
\usepackage{xcolor}
\usepackage{orcidlink}
\usepackage{graphicx}
\usepackage{ragged2e}

\newcommand{\nc}{\newcommand*}

\nc{\Eq}[1]{Eq.~\eqref{#1}}
\nc{\Fig}[1]{Fig.~\ref{#1}}
\nc{\Table}[1]{Table~\ref{#1}}
\nc{\Sec}[1]{Sec.~\ref{#1}}

\def\({\left(}
\def\){\right)}
\def\[{\left[}
\def\]{\right]}
\def\e{\begin{equation}}
\def\q{\end{equation}}
\def\m{\begin{eqnarray}}
\def\n{\end{eqnarray}}

\begin{document}

\title{Prospects for Detecting Post-merger Signals from Binary Neutron Stars with Future Gravitational-Wave Detectors}

\author{Jingwang Diao \orcidlink{0000-0002-3467-3102}}
\affiliation{School of Physics and Astronomy, Beijing Normal University, Beijing 100875, China}
\affiliation{Department of Physics, Faculty of Arts and Sciences, Beijing Normal University, Zhuhai 519087, China}

\author{Teng Zhang \orcidlink{0000-0001-5278-4220}}

\affiliation{School of Physics and Astronomy, and Institute for Gravitational Wave Astronomy, University of Birmingham, Edgbaston, Birmingham B15\,2TT, United Kingdom}

\author{Xingjiang Zhu \orcidlink{0000-0001-7049-6468}}
\thanks{Project supported by the National Key Research and Development Program of China (Grant No.~2023YFC2206704), the Fundamental Research Funds for the Central Universities, and the Supplemental Funds for Major Scientific Research Projects of Beijing Normal University (Zhuhai) (Grant No.~ZHPT2025001)}
\email{zhuxj@bnu.edu.cn}
\affiliation{Department of Physics, Faculty of Arts and Sciences, Beijing Normal University, Zhuhai 519087, China}
\affiliation{Institute for Frontiers in Astronomy and Astrophysics, Beijing Normal University, Beijing 102206, China}

\begin{abstract}
The post-merger (PM) phase of binary neutron star (BNS) mergers carries information about the merger remnant and the dense-matter equation of state, but its observational prospects remain uncertain. We compare two approaches to accessing this emission: the stochastic gravitational-wave background (SGWB) from the BNS population and direct PM-band detections of individually resolved mergers. Using a unified inspiral--merger--post-merger (IPM) waveform prescription and the same astrophysical population model for both calculations, we evaluate representative third-generation detector configurations involving ET, CE, and a dedicated kilohertz detector.
For the representative detector networks considered, the baseline IPM model predicts that the all-source, broadband BNS SGWB exceeds the adopted detection threshold across the local merger-rate interval. In contrast, the kilohertz PM-band SGWB remains below this threshold. For individual events, over $R_0=7.6$--$250~{\rm Gpc}^{-3}\,{\rm yr}^{-1}$, the baseline IPM model predicts approximately $5$--$174$ PM-band candidates per year for the dedicated kilohertz detector, whereas the numerical-relativity cases considered here yield approximately $0.07$--$7~{\rm yr}^{-1}$.
\end{abstract}
\maketitle

\section{Introduction}

The direct detection of gravitational waves (GWs) from compact binary coalescences has opened a new observational window on strong-field gravity, dense matter, and stellar evolution~\cite{GW150914,Abbott2017GW170817,Dietrich2021BNS}. In particular, GW170817 established binary neutron star (BNS) mergers as key multimessenger sources and demonstrated their value for probing strong-field gravity and dense-matter physics~\cite{Abbott2017GW170817,Abbott2017MMA}. Although the inspiral phase is now routinely modeled and targeted in searches with high precision, the post-merger (PM) phase remains far less explored, despite its potentially unique diagnostic power~\cite{BausweinJanka2012,Clark2016Postmerger,
TorresRivas2019Postmerger,Huez2026Kilohertz}.

The PM GW emission is produced by the merger remnant---a hypermassive or supramassive neutron star, or a promptly formed black hole, depending on the binary masses and the neutron star equation of state (EoS)~\cite{Bauswein2012EoS}. Its kilohertz-band spectrum encodes information about remnant dynamics, finite-density effects, and the EoS through characteristic oscillation frequencies and damping times \cite{Stergioulas2011,Bernuzzi2015Spectrum,Bose2018RadiusPopBNS}. Detecting PM signals from individual mergers is therefore a long-standing goal of GW astronomy. However, because PM emission is typically weaker than the inspiral signal and concentrated above $\sim 1~\mathrm{kHz}$, it is much more difficult to observe with current detectors \cite{Clark2016Postmerger,Chatziioannou2017PostMerger,LIGO170817pm_search,Abbott:2017xzg}.

Future third-generation (3G) ground-based detectors are expected to improve these prospects. Facilities such as the Einstein Telescope (ET)~\cite{Punturo2010,Maggiore2020ET} and Cosmic Explorer (CE)~\cite{Reitze2019CE,Evans2021CE} will extend the observable volume for compact binaries by orders of magnitude and substantially increase the number of detected BNS mergers. Several detector concepts have also been proposed specifically to improve sensitivity to kilohertz PM oscillations, including NEMO~\cite{Ackley2020NEMO} and a planned high-frequency detector concept in China. For definiteness, we refer to the latter in this paper as the ZhangHeng Telescope (ZT)~\cite{ZHTPhysRevX}, while noting that this is not yet an official name. Such enhanced high-frequency sensitivity may not only improve PM detectability but also enable the reconstruction of detailed remnant dynamics, including signatures associated with rotational instabilities in individual PM signals~\cite{Sasli:2023wpa}.

The BNS population is already expected to generate an astrophysical stochastic gravitational-wave background (SGWB), with the dominant contribution generally arising from the inspiral phase integrated over cosmic history \cite{Phinney2001,Rosado2011,Zhu2011}. More recently, attention has turned to the possible PM contribution in the kilohertz regime. In particular, Lehoucq et al.~\cite{Lehoucq2025Postmerger} showed that PM emission can significantly modify the high-frequency BNS background and may even dominate in this regime under favorable assumptions. These results highlight the importance of incorporating PM physics into population-level GW forecasts. They also make clear that detectability depends not only on the source spectrum itself, but also on detector sensitivity, network geometry, and the overlap reduction function (ORF) at high frequencies.

The stochastic and single-event aspects of BNS PM emission have been studied separately, but they are not usually compared using the same population and waveform assumptions. Such a comparison is useful for 3G detector networks that may combine broadband facilities, such as ET and CE, with high-frequency instruments such as ZT. This motivates an assessment, within a common framework, of the observational reach of the two channels and of how population assumptions, waveform modeling, and detector configurations shape their detectability.

Here we make this comparison for representative 3G detector configurations. We adopt the hybrid inspiral--merger--post-merger (IPM) waveform model of Ref.~\cite{Bose2018RadiusPopBNS,Lehoucq2025Postmerger}, use observationally motivated BNS merger-rate densities, and apply the same simulated population to the SGWB and single-event calculations. For the stochastic channel, we compute SGWB signal-to-noise ratios (SNRs) from the full waveform in both a broadband interval and a kilohertz PM-band interval. For the single-event channel, we estimate the annual number of mergers whose broadband signals and full-waveform PM-band signals are separately detectable. We also examine the effect of source removal, compare the analytic PM prescription with representative numerical relativity (NR) waveforms, and test two additional assumptions of the analytic PM model.

The remainder of this paper is organized as follows. In Sec.~\ref{sec:population}, we describe the BNS population and merger-rate model. In Sec.~\ref{sec:waveform}, we introduce the IPM waveform and the parameter prescription adopted in this work. In Sec.~\ref{sec:detectability}, we present the formalism for the SGWB and single-event detectability estimates, including the residual-background construction used after removing network-resolvable events. In Sec.~\ref{sec:results}, we summarize the main results for different detector networks, source-removal tests, and representative numerical-relativity waveforms. In Sec.~\ref{sec:conclusion}, we summarize our conclusions. Appendix~\ref{app:model_dependence} gives additional diagnostic checks of the tidal mapping and prompt-collapse assumptions.

\section{Astrophysical population and merger-rate model}
\label{sec:population}

The BNS population is described by a redshift-dependent merger-rate density~\cite{Zhu2013cbc,MadauDickinson2014,Dominik:2012pd,gwtc4pop} within a flat $\Lambda$CDM cosmology~\cite{Planck:2018vyg}. The same population model is used consistently for both observational channels considered here: the full cosmological population enters the SGWB calculation, whereas the subset of sufficiently loud mergers determines the single-event detection rates.

The source-frame BNS merger-rate density is written as $R(z)=R_0\,e(z)$, where $R_0$ is the local merger-rate density and $e(z)$ is a dimensionless evolution factor normalized such that $e(0)=1$~\cite{Zhu2013cbc}. Following the standard assumption that compact-binary formation broadly traces the cosmic star-formation history with a delay-time distribution, the evolution factor is defined as $e(z)=\dot{\rho}_{\ast,c}(z)/\dot{\rho}_{\ast,c}(0)$, where $\dot{\rho}_{\ast,c}(z)$ denotes the delay-time-convolved star-formation-rate density associated with compact binaries merging at redshift $z$. Following Ref.~\cite{Zhu2013cbc}, it is written as
\begin{equation}
\dot{\rho}_{\ast,c}(z)
=\int_{t_{\min}}^{t_{\ast}(z)}
\dot{\rho}_\ast(z_f)\,
\frac{dt_f}{dt_z}\,
p(t_d)\,dt_d .
\label{eq:conv}
\end{equation}
Here, $\dot{\rho}_\ast(z_f)$ is the cosmic star formation rate density evaluated at the formation redshift $z_f$, $t_d$ is the delay time between binary formation and merger, $p(t_d)$ is the delay-time probability distribution, and $t_{\min}$ is the minimum delay time. The upper limit $t_{\ast}(z)$ corresponds to the maximum delay time allowed for a system merging at redshift $z$. The formation redshift $z_f$ is determined by the lookback-time relation.
\begin{equation}
t_d =
\int_z^{z_f}
\frac{dz'}{(1+z')H(z')} .
\label{eq:tdelay}
\end{equation}
The factor $dt_f/dt_z=(1+z)/(1+z_f)$ converts the star-formation rate evaluated at the formation epoch into the corresponding contribution to the merger rate at redshift $z$.

For the cosmic star-formation rate density, we adopt the Madau--Dickinson~\cite{MadauDickinson2014} parametrization,
\begin{equation}
\dot{\rho}_\ast(z)
=
0.015\,
\frac{(1+z)^{2.7}}
{1+\left[(1+z)/2.9\right]^{5.6}}
\;M_\odot\,\mathrm{yr}^{-1}\,\mathrm{Mpc}^{-3},
\label{eq:md14}
\end{equation}
The delay-time distribution is assumed to follow the commonly used power-law form
$p(t_d)\propto t_d^{-1}$ for $t_d \ge t_{\min}$~\cite{Dominik:2012pd},
with a fiducial minimum delay time of $t_{\min}=20~\mathrm{Myr}$. 

The overall normalization is set by the local BNS merger-rate density inferred from recent LVK population analyses. Unless otherwise stated, this work adopts the fiducial GWTC-4.0 range, $R_0=7.6$--$250~\mathrm{Gpc}^{-3}\,\mathrm{yr}^{-1}$, consistent with the latest population analysis~\cite{gwtc4pop}. For the final tabulated SGWB and event-count comparisons, the low-rate endpoint is obtained by linearly rescaling the results computed at $R_0=250~\mathrm{Gpc}^{-3}\,\mathrm{yr}^{-1}$. This convention keeps the rate dependence explicit and avoids interpreting finite-catalog fluctuations from separate low-rate realizations as physical differences.

Given $R(z)$, the differential number of mergers per unit redshift and observer time is
\begin{equation}
\frac{d\dot{N}}{dz}
=
\frac{R(z)}{1+z}\,\frac{dV_c}{dz},
\label{eq:dNdz}
\end{equation}
where $dV_c/dz$ is the comoving volume element, and the factor $(1+z)^{-1}$ accounts for cosmological time dilation between the source and observer frames. The total observed merger rate is then given by
\begin{equation}
\dot{N}_{\rm tot}
=
\int_0^{z_{\max}}
\frac{R(z)}{1+z}\,
\frac{dV_c}{dz}\,dz.
\label{eq:Ntot}
\end{equation}
In the Monte Carlo calculations, we truncate the redshift integration and source sampling at $z_{\max}=10$, which is sufficient for the adopted BNS merger-rate model and the detector bands considered here.

Throughout this paper, we assume a flat $\Lambda$CDM cosmology with
$H_0=67.4~\mathrm{km\,s^{-1}\,Mpc^{-1}}$, $\Omega_m=0.315$, and
$\Omega_\Lambda=0.685$, consistent with the Planck 2018 results~\cite{Planck:2018vyg}.

\section{Inspiral--merger--post-merger waveform model}
\label{sec:waveform}

\begin{figure}[t]
    \centering
    \includegraphics[width=\linewidth]{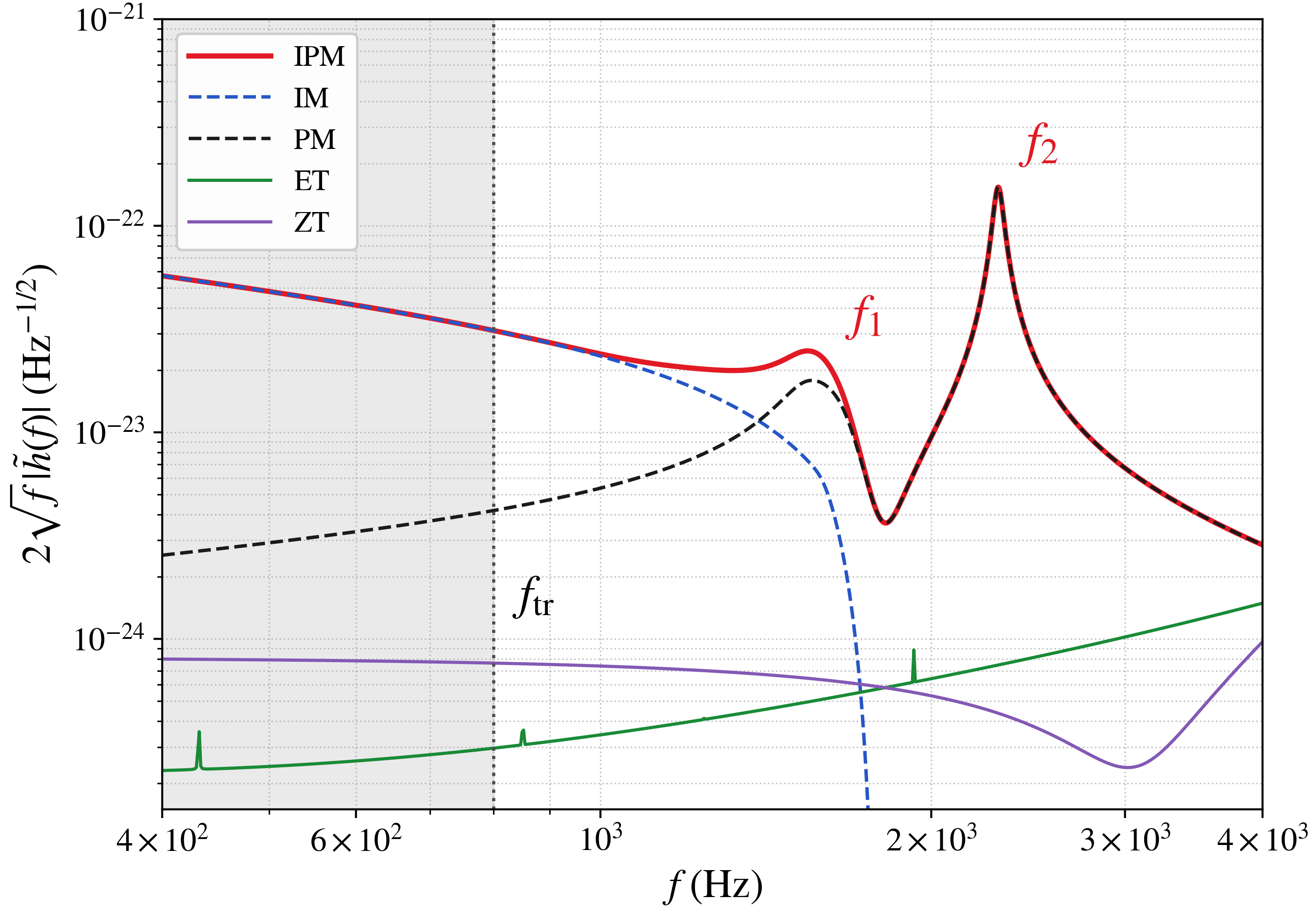}
    \caption{\justifying Representative baseline IPM amplitude spectrum, together with the ET and ZT amplitude sensitivity curves adopted from Ref.~\cite{ZHTPhysRevX}. The illustrative source is an equal-mass binary with $m_1=m_2=1.25\,M_\odot$ and $d_L=50~\mathrm{Mpc}$. The red solid, blue dashed, and black dashed curves represent the IPM, inspiral-merger (IM), and PM spectra, respectively. 
    }
    \label{fig:ipm_example}
\end{figure}
 
For each BNS event, we follow the IPM waveform prescription of Ref.~\cite{Lehoucq2025Postmerger}. The inspiral--merger (IM) component is generated with the \texttt{IMRPhenomD\_NRTidalv2} approximant~\cite{Khan2016IMRPhenomD,Dietrich2019NRTidalv2}, evaluated through the \texttt{PyCBC} interface~\cite{PyCBCSoftware}.

The PM signal is modeled using the analytic prescription introduced in Ref.~\cite{Bose2018RadiusPopBNS} and adopted in Ref.~\cite{Lehoucq2025Postmerger}. The explicit time-domain expression for the $+$ polarization is,
\begin{align}
h_{+}^{\rm PM}(t) &=
h_0\Bigg[
\alpha\,e^{-t/\tau_1}
\Big(
\sin 2\pi f_1 t
+\sin 2\pi (f_1-f_{1\epsilon}) t
\nonumber\\&\qquad
+\sin 2\pi (f_1+f_{1\epsilon}) t
\Big)
\nonumber\\
&\qquad
+e^{-t/\tau_2}
\sin\!\Big(2\pi f_2 t + 2\pi\gamma_2 t^2 + 2\pi\xi_2 t^3 + \pi\beta_2\Big)
\Bigg],
\label{eq:pm_time}
\end{align}
where $t=0$ denotes the merger time. Following Ref.~\cite{Lehoucq2025Postmerger}, the overall amplitude scale is $h_0=G\,\mathcal{M}_c/(2\,d_L\,c^2)$. Here, $\mathcal{M}_c$ is the chirp mass and $d_L$ is the luminosity distance. In Eq.~\eqref{eq:pm_time}, $\alpha$ sets the relative amplitude of the low-frequency PM component; $\tau_1$ and $\tau_2$ are the damping times of the $f_1$ and $f_2$ components, respectively; $f_1$ and $f_2$ denote the corresponding characteristic PM frequencies; $f_{1\epsilon}$ controls the splitting around $f_1$; and $\beta_2$, $\gamma_2$, and $\xi_2$ describe the phase offset and higher-order phase evolution of the dominant component. Following Refs.~\cite{Bose2018RadiusPopBNS,Lehoucq2025Postmerger}, we fix the dimensionless and phase-evolution parameters to their fiducial values and determine the characteristic frequencies and damping times using the prescription calibrated in those works. The cross polarization $h_{\times}^{\rm PM}(t)$ is constructed from the same analytic PM ansatz, with a quadrature ($\pi/2$) phase shift relative to $h_{+}^{\rm PM}(t)$.

To connect the inspiral parameters to the PM spectrum without adopting a specific EoS, we follow the EoS-agnostic procedure of Ref.~\cite{Lehoucq2025Postmerger}. For each equal-mass binary, the tidal deformability is assigned from the empirical $\mathcal{M}_c$--$\Lambda$ relation \cite{Altiparmak2022MchirpLambda}. The quantities $f_1$, $f_2$, $\tau_1$, and $\tau_2$ are then determined from the quasi-universal relations summarized in Refs.~\cite{Bose2018RadiusPopBNS,Lehoucq2025Postmerger}. We use the central $\mathcal{M}_c$--$\Lambda$ relation as a baseline phenomenological mapping within the empirical range. This choice does not imply that the adopted relation represents a preferred EoS; it simply provides a definite mapping from $\mathcal M_c$ to the tidal and PM parameters required by the analytic waveform. The PM contribution is Fourier transformed numerically and combined with the IM waveform following Ref.~\cite{Lehoucq2025Postmerger}. The transition frequency is set to $f_{\rm tr}=f_1/2$. Below $f_{\rm tr}$, the hybrid waveform is given by the IM waveform, because the PM contribution is negligible in this region. Above $f_{\rm tr}$, the complex IM and PM strains are summed. To ensure phase continuity between the two components, their phases are matched at $f_{\rm tr}$. Fig.~\ref{fig:ipm_example} illustrates the resulting IPM amplitude spectrum together with the individual IM and PM contributions.

We use this IPM waveform as the baseline model because it is sufficiently fast for Monte Carlo calculations over a cosmological BNS population. The analytic PM prescription describes the oscillatory emission of a neutron star remnant. For sufficiently massive systems, prompt collapse to a black hole can strongly suppress or eliminate this remnant-dominated PM component, leaving a qualitatively different and typically weaker high-frequency signal. We do not impose an EoS-dependent prompt-collapse threshold in the baseline population. The high-mass tail may therefore overestimate the PM contribution from prompt-collapse systems.

As a check on waveform-model dependence, we also repeat the calculation with three representative equal-mass, nonspinning NR waveforms from the CoRe database, labeled by their equations of state: SLy, H4, and MS1b \cite{CoReDietrich2018}. We choose this compact set because the simulations have similar total masses and binary configurations while spanning a soft-to-stiff EoS sequence with systematically different radii and tidal deformabilities. The comparison therefore probes whether the main PM-band conclusions depend on the analytic waveform prescription or on a single assumed EoS, without introducing additional variations from spin or mass ratio.

For the baseline IPM population, each simulated source is assigned a redshift drawn from the population model of Sec.~\ref{sec:population}, namely from $d\dot N/dz \propto R(z)(dV_c/dz)/(1+z)$, and the corresponding luminosity distance is computed from the adopted cosmology. Under the equal-mass assumption, the intrinsic mass is specified by a single component mass $m$, drawn from the turn-on power-law (TOP) model~\cite{You2025BirthMass}, with $m_1=m_2\equiv m$; the chirp mass $\mathcal{M}_c$ then follows accordingly\footnote{In the NR comparison, the component masses are fixed to those of the selected simulations rather than drawn from the TOP model; the redshift and distance prescriptions are otherwise unchanged.}. The EoS-agnostic prescription described above is used to assign $\Lambda$ from $\mathcal{M}_c$ and hence to determine the PM quantities entering Eq.~\eqref{eq:pm_time}. The extrinsic parameters are sampled assuming isotropic sky positions and source orientations: right ascension is drawn uniformly in $[0,2\pi)$, declination uniformly in $\sin(\mathrm{dec})\in[-1,1]$, polarization angle $\psi$ uniformly in $[0,\pi)$, and inclination uniformly in $\cos\iota\in[-1,1]$. In this baseline study, the coalescence phase is fixed to $\phi_c=0$\footnote{We verified that randomizing the coalescence phase does not change the event SNRs.}. For convenience, Table~\ref{tab:parameter_sampling} summarizes the main population, geometry, and waveform parameters used in this work.

These choices keep the SGWB and single-event calculations tied to the same population model. The equal-mass assumption follows the simplified IPM population treatment adopted in Ref.~\cite{Lehoucq2025Postmerger}. It removes the degeneracy between chirp mass and tidal deformability in the EoS-agnostic mapping and allows a single value of $\Lambda$ to be associated with each sampled $\mathcal{M}_c$. This is a modeling simplification, not a statement that the astrophysical BNS population is exactly equal mass. If unequal masses were included, the same chirp mass would correspond to different component masses and, in general, to different $(\Lambda_1,\Lambda_2)$. The PM mapping would therefore require an EoS ensemble or an additional prescription for the mass ratio and individual tidal deformabilities. These simplifications may affect the quantitative PM-sensitive forecasts. 

\begin{table*}[t]
\centering
\caption{\justifying Primary population, geometric, and IPM waveform parameters used in this work. Quantities are fixed, sampled from the listed distributions, or derived from sampled variables.}
\label{tab:parameter_sampling}
\begin{tabular}{llll}
\hline
\textbf{Category} & \textbf{Parameter} & \textbf{Sampling / prescription} & \textbf{Notes} \\
\hline
Population & Redshift $z$ & Drawn from $d\dot N/dz \propto R(z)\,(dV_c/dz)/(1+z)$ & Source-frame merger-rate model \\
Population & Local merger rate $R_0$ & Fixed scenario value & Baseline values: $7.6$ and $250~\mathrm{Gpc}^{-3}\,\mathrm{yr}^{-1}$ \\
Population & Component mass $m$ & Sampled from the TOP model & Equal-mass assumption: $m_1=m_2\equiv m$ \\
Geometry & $(\mathrm{ra},\mathrm{dec},\psi,\iota)$ & Isotropic distributions & Used in the single-event calculation \\
Geometry & Coalescence phase $\phi_c$ & Fixed to $0$ & Baseline simplification \\
Derived & Luminosity distance $d_L$ & Derived from $z$ & Flat $\Lambda$CDM cosmology \\
Derived & Chirp mass $\mathcal{M}_c$ & Derived from $m$ & Equal-mass case \\
Derived & Tidal deformability $\Lambda$ & Derived from $\mathcal{M}_c$ & EoS-agnostic prescription \\
Derived & PM quantities $(f_1,f_2,\tau_1,\tau_2,h_0)$ & Derived from $\mathcal{M}_c$, $\Lambda$, and $d_L$ & Analytic PM prescription of Refs.~\cite{Bose2018RadiusPopBNS,Lehoucq2025Postmerger} \\
\hline
\end{tabular}
\end{table*}

\section{Detectability in two observational channels}
\label{sec:detectability}

\begin{figure}[t]
    \centering
    \includegraphics[width=\linewidth]{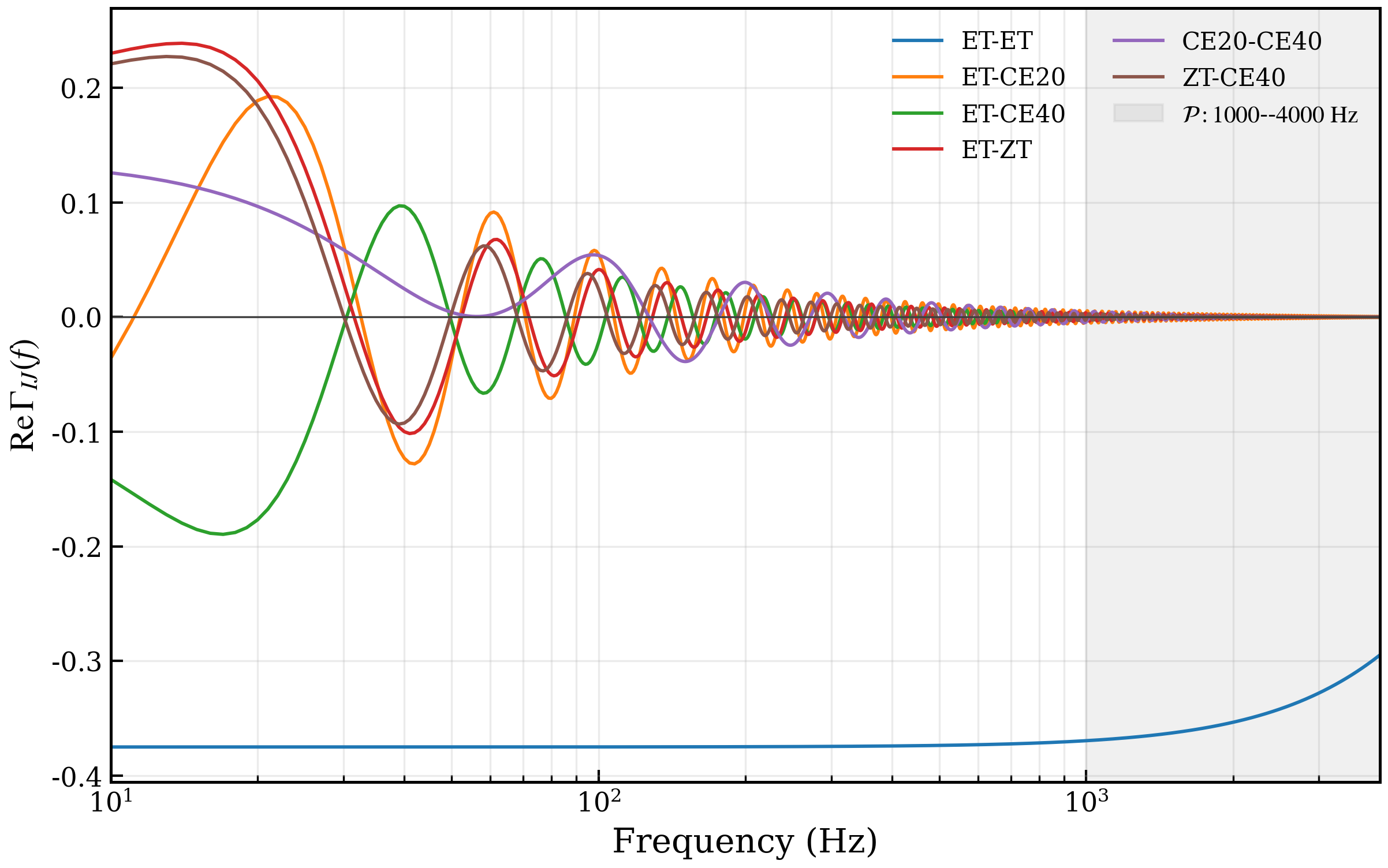}
    \caption{\justifying Real parts of representative frequency-dependent overlap reduction functions for selected detector baselines. The imaginary parts are omitted for clarity. The shaded region indicates the adopted PM-band interval of 1000--4000 Hz.}
    \label{fig:orf_selected}
\end{figure}

We use two detectability measures: one for the SGWB generated by the simulated BNS population and another for individually resolvable mergers.

For the stochastic channel, the relevant observable is the dimensionless GW energy-density spectrum \cite{AllenRomano1999,Phinney2001},
\begin{equation}
\Omega_{\rm GW}(f)
\equiv
\frac{1}{\rho_c}\frac{d\rho_{\rm GW}}{d\ln f},
\qquad
\rho_c=\frac{3H_0^2c^2}{8\pi G},
\label{eq:OmegaGW_def}
\end{equation}
where $\rho_c$ denotes the critical energy density of the Universe. Rather than evaluating $\Omega_{\rm GW}(f)$ using a purely analytical population integral, we construct it by stacking the simulated waveform catalog over an observation time $T$. For a population of independent sources \cite{Lehoucq2025Postmerger},
\begin{equation}
\Omega_{\rm GW}(f)
=
\frac{4\pi^{2}}{3H_{0}^{2}\,T}\,f^{3}
\sum_{i=1}^{N}
\left(
\left|\tilde h^{+}_{i}(f)\right|^{2}
+
\left|\tilde h^{\times}_{i}(f)\right|^{2}
\right),
\label{eq:Omega_stack_final}
\end{equation}
where $\tilde h^{+}_{i}(f)$ and $\tilde h^{\times}_{i}(f)$ are the Fourier amplitudes of the two polarizations for the $i$th source, and $N$ is the total number of mergers occurring during the observing time. In this work, we set $T=1~\mathrm{yr}$. Because event times are random, cross terms average out, and the background is given by the incoherent sum of the individual source spectra \cite{AllenRomano1999,Phinney2001}. In the baseline forecast, all simulated mergers during the observing time are included in the stacking analysis. 

The detectability of the SGWB is quantified using the standard cross-correlation SNR. For a detector pair $(I,J)$, we write \cite{AllenRomano1999}
\begin{equation}
\rho_{IJ}^{2} = \frac{9H_0^4}{50\pi^4}\, T
\int_{f_{\min}}^{f_{\max}} df\,
\frac{|\Gamma_{IJ}(f)|^{2}\,\Omega_{\rm GW}^{2}(f)}
     {f^{6}P_I(f)P_J(f)} ,
\label{eq:snr_pair_omega}
\end{equation}
where $P_I(f)$ and $P_J(f)$ are the one-sided noise power spectral densities of the two detectors, and $\Gamma_{IJ}(f)$ is the overlap reduction function (ORF) of the corresponding detector baseline. For an isotropic, unpolarized SGWB, the ORF is defined as
\cite{AllenRomano1999,Flanagan1993}
\begin{equation}
\begin{aligned}
\Gamma_{IJ}(f)
={}&\frac{5}{8\pi}
\sum_{A=+,\times}
\int_{S^2} d\hat{\boldsymbol{\Omega}}\,
F_I^A(f,\hat{\boldsymbol{\Omega}})
F_J^{A*}(f,\hat{\boldsymbol{\Omega}})
\\
&\times
\exp\!\left[
\frac{2\pi i f}{c}
\hat{\boldsymbol{\Omega}}\cdot
\left(\boldsymbol{x}_I-\boldsymbol{x}_J\right)
\right],
\end{aligned}
\label{eq:orf_definition}
\end{equation}
where $\boldsymbol{x}_I$ and $\boldsymbol{x}_J$ are the detector
locations, $\hat{\boldsymbol{\Omega}}$ denotes the GW propagation
direction, and $A=+,\times$ labels the two tensor polarizations. Instead of adopting the long-wavelength approximation, we evaluate the antenna responses in Eq.~\eqref{eq:orf_definition} using the finite-arm formalism of Ref.~\cite{Essick2017FiniteArm}. The frequency dependence of the ORF therefore arises from both the propagation phase across the detector baseline and the finite-arm transfer functions. The resulting ORF is generally complex, and the full $|\Gamma_{IJ}(f)|^2$ is retained in the SNR calculation. The normalization is chosen such that a co-located and co-aligned right-angle interferometer pair has $\Gamma_{IJ}(0)=1$.

For a detector network, the baselines are combined in quadrature under the idealized assumption that detector noises are mutually uncorrelated and that covariance between baseline estimators can be neglected:
\begin{equation}
\rho_{\rm net}
=
\left(
\sum_{I<J}\rho_{IJ}^{2}
\right)^{1/2}.
\label{eq:rho_net}
\end{equation}

When assessing whether a stochastic signal is detectable, we adopt $\rho_{\rm SGWB}=3$ as the reference cross-correlation SNR threshold \cite{RomanoCornish2017}. Values below this threshold are therefore regarded as sub-threshold or marginal prospects rather than robust detections.

The strain-noise sensitivity curves adopted for ZT, ET, CE20, and CE40 are taken from Ref.~\cite{ZHTPhysRevX}. The ORFs are evaluated for the adopted detector sites, orientations, and arm lengths. For ET and CE, we use the fiducial geometries summarized in Table~2 of Ref.~\cite{Gupta2024}. The ET triangular design is modeled as three $60^\circ$ Michelson interferometers at the vertices of a 10-km equilateral triangle, providing three internal baselines for cross-correlation analyses \cite{Punturo2010,Branchesi:2023mws}. The CE20 and CE40 labels denote the 20-km and 40-km L-shaped Cosmic Explorer configurations used in that network study \cite{Reitze2019CE,Evans2021CE}. For ET+CE20+CE40+ZT, we also include a fiducial right-angle ZT detector located at latitude $34.5^\circ$ and longitude $114.3^\circ$ \cite{Li2019located_zht}. Figure~\ref{fig:orf_selected} shows the real parts of representative ORFs\footnote{For clarity, only a subset of baselines is shown. For ET we plot one representative internal ET--ET baseline rather than all three. Likewise, only one representative baseline is displayed for each ET--CE and ET--ZT pairing.}. To illustrate the impact of geometric suppression, we also consider an idealized colocated ZT benchmark for which the ORF is artificially set to unity. This ORF $=1$ case is only an upper benchmark and does not represent a realized detector pair or full network.

With this detector setup specified, we construct a residual background to assess the effect of source removal. Events are excluded from the stack if they are broadband-resolvable by the ET+CE20+CE40+ZT single-event network, where the broadband network SNR is computed as the quadrature sum of the corresponding single-detector broadband SNRs. We remove events satisfying $\rho_{\rm ev,net}^{\mathcal{B}}\ge 12$, following ET forecasting studies in which an SNR threshold of 12 is used to define resolved sources before constructing residual compact-binary backgrounds~\cite{Branchesi:2023mws}. Here, the criterion is a uniform proxy for identifying the loudest broadband-resolvable sources in the detector network, and those sources are removed from the stack under idealized assumptions. The resulting residual background is therefore a source-removal test, not a realistic subtraction forecast: it does not include waveform-subtraction errors, parameter-estimation uncertainties, overlapping signals, or iterative subtraction. We use it only to determine whether removing the loudest broadband events changes the relative detectability of the broadband and PM-band SGWB components.

To distinguish broadband detectability from the high-frequency PM-band content, two frequency intervals are used throughout this work:
\begin{equation}
\mathcal{B}=[1,4000]~\mathrm{Hz}, \qquad
\mathcal{P}=[1000,4000]~\mathrm{Hz}.
\end{equation}
We refer to $\mathcal{B}$ as the broadband interval and to $\mathcal{P}$ as the PM-band interval. The PM-band interval is selected to isolate the frequency range most directly associated with remnant oscillations. The GW170817 post-merger search used $1024$--$4096~\mathrm{Hz}$ for its short-duration analysis and reported short-duration signal upper limits over $1$--$4~\mathrm{kHz}$, while detector-design studies for neutron-star physics identify improved sensitivity from $1$ to $4~\mathrm{kHz}$ as central to measuring BNS post-merger dynamics~\cite{Abbott:2017xzg,Miao:2017iqq}. For both the SGWB and single-event channels, we calculate SNRs in both intervals using the complete IPM waveform, thereby maintaining a common population and waveform framework. The PM-band SNR therefore includes any IM contribution and IM--PM interference within this interval and characterizes high-frequency detectability rather than an isolated PM component. The corresponding SGWB SNRs are denoted by $\rho_{\rm SGWB}^{\mathcal{B}}$ and $\rho_{\rm SGWB}^{\mathcal{P}}$.

For the single-event channel, detectability is quantified using the standard matched-filter SNR~\cite{Flanagan1997snr},
\begin{equation}
\rho^{2}
=
4\int_{f_{\min}}^{f_{\max}}
\frac{|\tilde h(f)|^{2}}{P(f)}\,df .
\label{eq:single_snr}
\end{equation}
These SNRs assume perfect waveform matching between the signal and the filtering template and include no penalty for template mismatch.

Using the IPM waveform model of Sec.~\ref{sec:waveform}, we evaluate two event-level SNRs for each simulated source: $\rho_{\rm ev}^{\mathcal{B}}$ over the broadband interval and $\rho_{\rm ev}^{\mathcal{P}}$ over the PM-band interval. Following the same forecast-level convention, the annual rate of events satisfying $\rho_{\rm ev}^{\mathcal{B}}\ge 12$ is denoted by $\dot N_{\mathcal{B}}$. The PM-band candidate subset is denoted by $\dot N_{\mathcal{P}}$ and requires both $\rho_{\rm ev}^{\mathcal{B}}\ge 12$ and $\rho_{\rm ev}^{\mathcal{P}}\ge 12$. Thus, $\dot N_{\mathcal{B}}$ measures the rate of broadband-detectable BNS mergers, whereas $\dot N_{\mathcal{P}}$ measures the rate of such mergers whose full-waveform high-frequency signal is detectable in the PM band.

We use the same SNR threshold for the broadband and PM-band single-event rates. This conservative convention keeps $\dot N_{\mathcal{B}}$ and $\dot N_{\mathcal{P}}$ directly comparable and matches the source-removal threshold used in the residual-background calculation. It is not an optimized threshold for a targeted PM follow-up search. Dedicated PM detector-design studies have adopted lower thresholds, for example $\rho_{\rm PM}>5$, to estimate more optimistic PM detection rates~\cite{ZHTPhysRevX}. A lower threshold would increase the predicted PM-band candidate rates.

\section{Results and Discussion}
\label{sec:results}

\begin{figure}[t]
    \centering
    \includegraphics[width=1.0\linewidth]{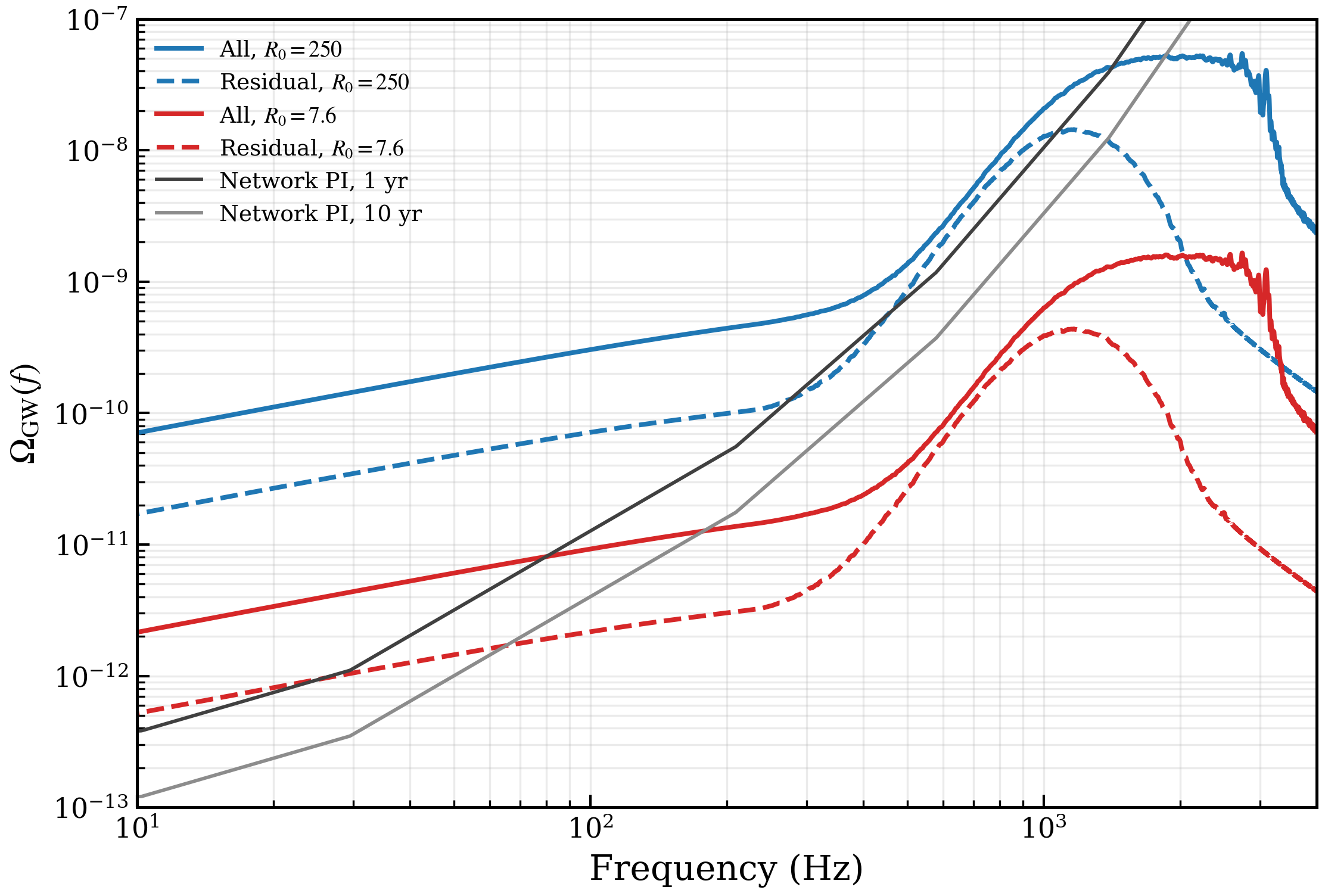}
    \caption{\justifying Baseline IPM predictions are shown for the BNS stochastic-background spectrum at $R_0=250~\mathrm{Gpc}^{-3}\,\mathrm{yr}^{-1}$ and at the scaled low-rate endpoint $R_0=7.6~\mathrm{Gpc}^{-3}\,\mathrm{yr}^{-1}$. Solid curves denote the all-source background, whereas dashed curves denote the residual background after ideal removal of events with $\rho_{\rm ev,net}^{\mathcal{B}}\ge 12$. The PI curves are shown for ET+CE20+CE40+ZT and use the conventional expected-SNR normalization $\rho_{\rm SGWB}=1$, with the observation times indicated in the legend.}
    \label{fig:sgwb_spectra}
\end{figure}

\begin{table*}[t]
\centering
\caption{\justifying SGWB cross-correlation SNRs for the baseline IPM background. Entries report $\rho_{\rm SGWB}^{\mathcal{B}}$, with $\rho_{\rm SGWB}^{\mathcal{P}}$ shown in parentheses. ``All-source'' denotes the background from the full simulated population, whereas ``Residual'' denotes the background after removing events with $\rho_{\rm ev,net}^{\mathcal{B}}\ge 12$ in the ET+CE20+CE40+ZT single-event network. ZT (ORF$=1$) denotes an idealized co-located benchmark. The nearly identical PM-band SNRs of the three detector-network configurations reflect the dominance of the internal ET--ET baselines at kilohertz frequencies, with the inter-site baselines contributing negligibly at the quoted precision.}
\label{tab:sgwb_snr_baseline}
\begin{tabular}{llcc}
\hline
\textbf{Network} & \textbf{Background} & \textbf{$R_0=7.6$} & \textbf{$R_0=250$} \\
\hline
ET                  & All-source & 4.26 (0.0212) & 140 (0.697) \\
ET                  & Residual   & 1.03 (0.00972) & 34.0 (0.320) \\
ET+CE20+CE40        & All-source & 5.04 (0.0212) & 166 (0.697) \\
ET+CE20+CE40        & Residual   & 1.22 (0.00972) & 40.2 (0.320) \\
ET+CE20+CE40+ZT     & All-source & 5.40 (0.0212) & 178 (0.697) \\
ET+CE20+CE40+ZT     & Residual   & 1.31 (0.00972) & 43.0 (0.320) \\
ZT (ORF$=1$)        & All-source & 2.49 (0.0680) & 81.8 (2.24) \\
ZT (ORF$=1$)       & Residual   & 0.60 (0.0142) & 19.7 (0.466) \\
\hline
\end{tabular}
\end{table*}

\begin{table*}[t]
\centering
\caption{\justifying Predicted annual rates of individually detectable BNS events for the baseline IPM waveform model. Entries list $\dot N_{\mathcal{B}}$, with the PM-band candidate rate $\dot N_{\mathcal{P}}$ shown in parentheses, in units of $\mathrm{yr}^{-1}$. }
\label{tab:single_event_counts_ipm}
\begin{tabular}{lcc}
\hline
\textbf{Detector} & \textbf{$R_0=7.6$} & \textbf{$R_0=250$} \\
\hline
ET   & 862.90 (0.33)   & 28385 (11) \\
ZT   & 79.28 (5.29)    & 2608 (174) \\
CE20 & 101.20 (0.58)  & 3329 (19) \\
CE40 & 2377.77 (0.21) & 78216 (7) \\
\hline
\end{tabular}
\end{table*}

\subsection{Stochastic-background detectability}

Fig.~\ref{fig:sgwb_spectra} shows the baseline and residual SGWB spectra from the IPM model, together with representative power-law integrated (PI) sensitivity curves~\cite{Thrane:2013oya}. The solid curves are obtained by stacking all simulated events, whereas the dashed curves exclude sources satisfying the ET+CE20+CE40+ZT broadband network criterion, $\rho_{\rm ev,net}^{\mathcal{B}}\ge 12$. The two local-rate endpoints differ primarily in their overall normalization, following the scaling convention in Sec.~\ref{sec:population}. As in Ref.~\cite{Lehoucq2025Postmerger}, the spectrum contains a lower-frequency inspiral--merger contribution and a broad kilohertz feature associated with PM emission.

Table~\ref{tab:sgwb_snr_baseline} gives the corresponding SNRs. Over $\mathcal{B}$, the predicted all-source background for ET yields $\rho_{\rm SGWB}^{\mathcal{B}}=4.26$ at the scaled low-rate endpoint, $R_0=7.6~\mathrm{Gpc}^{-3}\,\mathrm{yr}^{-1}$, and $140$ at $R_0=250~\mathrm{Gpc}^{-3}\,\mathrm{yr}^{-1}$. The ET+CE20+CE40 and ET+CE20+CE40+ZT networks yield $5.04$ and $5.40$, respectively, at the low-rate endpoint and remain above the adopted threshold across the rate interval. The PM-band SNR is much smaller: ET gives $\rho_{\rm SGWB}^{\mathcal{P}}=0.0212$ and $0.697$ at the two rate endpoints, and the larger detector networks yield nearly identical values at the precision shown. Thus, for the detector-network configurations with the adopted geometries, the all-source PM-band SNR remains below unity throughout the adopted local-rate interval.

Eq.~\eqref{eq:snr_pair_omega} explains the large difference between the broadband and PM-band SNRs. First, the explicit $f^{-6}$ factor strongly down-weights kilohertz emission relative to the lower-frequency inspiral contribution. Second, the detector-noise product $P_I(f)P_J(f)$ provides additional suppression in the kilohertz band for the adopted sensitivity curves. Third, the ORFs of geographically separated detectors are strongly reduced at high frequencies. The propagation phase in Eq.~\eqref{eq:orf_definition} is controlled by $2\pi fD_{IJ}/c$, where $D_{IJ}=|\boldsymbol{x}_I-\boldsymbol{x}_J|$; increasing either the frequency or the detector separation produces more rapid phase variation across the sky and, consequently, stronger cancellation in the angular integral. The finite-arm transfer functions introduce additional frequency dependence when $fL_I/c$ is no longer negligible. As a result, the internal ET--ET baselines retain an appreciable PM-band response, whereas the inter-site ET--CE, CE--CE, and ZT baselines are strongly suppressed. Because all three network configurations contain the same three internal ET baselines, these baselines dominate their PM-band SNRs.

This geometric effect also limits the stochastic benefit of adding ZT. Relative to ET+CE20+CE40, including ZT increases $\rho_{\rm SGWB}^{\mathcal{B}}$ from $5.04$ to $5.40$ at the scaled low-rate endpoint and from $166$ to $178$ at the high-rate endpoint, whereas the change in $\rho_{\rm SGWB}^{\mathcal{P}}$ is negligible at the precision shown in Table~\ref{tab:sgwb_snr_baseline}. The idealized colocated ZT benchmark with $\Gamma_{IJ}=1$ instead gives $\rho_{\rm SGWB}^{\mathcal{P}}=0.0680$ and $2.24$ at the two endpoints. Even in this optimistic benchmark, the PM-band SNR remains below the adopted stochastic threshold, illustrating that favorable kilohertz strain sensitivity alone is insufficient for a stochastic measurement without a correspondingly favorable high-frequency ORF.

These forecasts depend on the adopted population and PM waveform model. Longer damping times can strengthen the PM background, as illustrated in Fig.~2 of Ref.~\cite{Lehoucq2025Postmerger}; our reference damping time of $50~\mathrm{ms}$ for the dominant PM component matches their fiducial choice. For a fixed background spectrum and detector configuration, longer observations also improve detectability through $\rho_{\rm SGWB}\propto\sqrt{T}$.

To assess the temporal overlap of PM signals, we estimate their effective duty cycle as~\cite{Zhu2013cbc}
\begin{equation}
D=\int_0^{z_{\max}}\Delta\tau(z)
\frac{d\dot N}{dz}\,dz ,
\label{eq:pm_duty_cycle}
\end{equation}
where $d\dot N/dz$ is the observer-frame differential merger rate, and $\Delta\tau(z)$ is the population-averaged effective PM duration in the observer frame, including cosmological time dilation. Evaluating this expression for the full simulated population gives $D=7.7\times10^{-3}$ at $R_0=250~\mathrm{Gpc}^{-3}\,\mathrm{yr}^{-1}$ and $D=2.3\times10^{-4}$ at $R_0=7.6~\mathrm{Gpc}^{-3}\,\mathrm{yr}^{-1}$. Both values are far below unity, indicating sparse, intermittent PM emission rather than continuous overlap within the adopted model.

Source removal substantially reduces the BNS background but leaves a non-negligible residual. For the ET+CE20+CE40+ZT network, the broadband SNR decreases from $178$ to $43.0$ at the upper merger-rate limit, whereas the PM-band SNR decreases from $0.697$ to $0.320$ and remains well below the adopted detection threshold. Although the residual PM component is not detectable under this criterion, the residual broadband background may constitute a relevant astrophysical foreground in searches for weaker cosmological SGWBs. Around $100~\mathrm{Hz}$, the residual spectrum spans approximately $\Omega_{\rm GW}\sim10^{-12}$--$10^{-10}$ across the adopted merger-rate interval, overlapping the range of foreground-limited sensitivities estimated for cosmological-SGWB searches with third-generation detectors~\cite{Zhong2023Cosmological}. This comparison highlights the importance of realistic foreground subtraction and residual modeling for future cosmological SGWB searches.

\subsection{Single-event detectability}

Table~\ref{tab:single_event_counts_ipm} reports the single-event rates for the baseline IPM model, where $\dot N_{\mathcal{B}}$ denotes the broadband-detectable rate and $\dot N_{\mathcal{P}}$ in parentheses denotes the PM-band candidate rate. Thus, a BNS merger may satisfy the broadband criterion while its high-frequency signal remains below the PM-band threshold. We adopt $\rho\ge 12$ for both $\rho_{\rm ev}^{\mathcal{B}}$ and $\rho_{\rm ev}^{\mathcal{P}}$, consistent with the ET design-forecast threshold~\cite{Branchesi:2023mws} and the source-removal criterion used above.

At the scaled low-rate endpoint, CE40 and ET yield the largest broadband-detectable rates, whereas ZT has a substantially smaller broadband rate but the largest PM-band candidate rate. In the baseline IPM calculation, $\dot N_{\mathcal{P}}$ for ZT is roughly an order of magnitude larger than those for ET and CE. The same relative behavior occurs at the upper merger-rate endpoint, because the event rates scale linearly with the adopted local merger-rate density.

The different detector ranking reflects the frequency ranges emphasized by the two observables. Broadband detectability is dominated by the stronger inspiral--merger signal at lower frequencies, favoring ET and CE40 and yielding a larger overall BNS sample. In contrast, $\dot N_{\mathcal{P}}$ selects the much smaller subset of events with sufficiently strong kilohertz emission, for which the enhanced high-frequency sensitivity of ZT becomes more important. This behavior is consistent with Ref.~\cite{ZHTPhysRevX}, which similarly found substantial gains in individual post-merger observations from a high-frequency-optimized detector.

\subsection{Impact of waveform modeling}

\begin{figure}[t]
    \centering
    \includegraphics[width=1.0\linewidth]{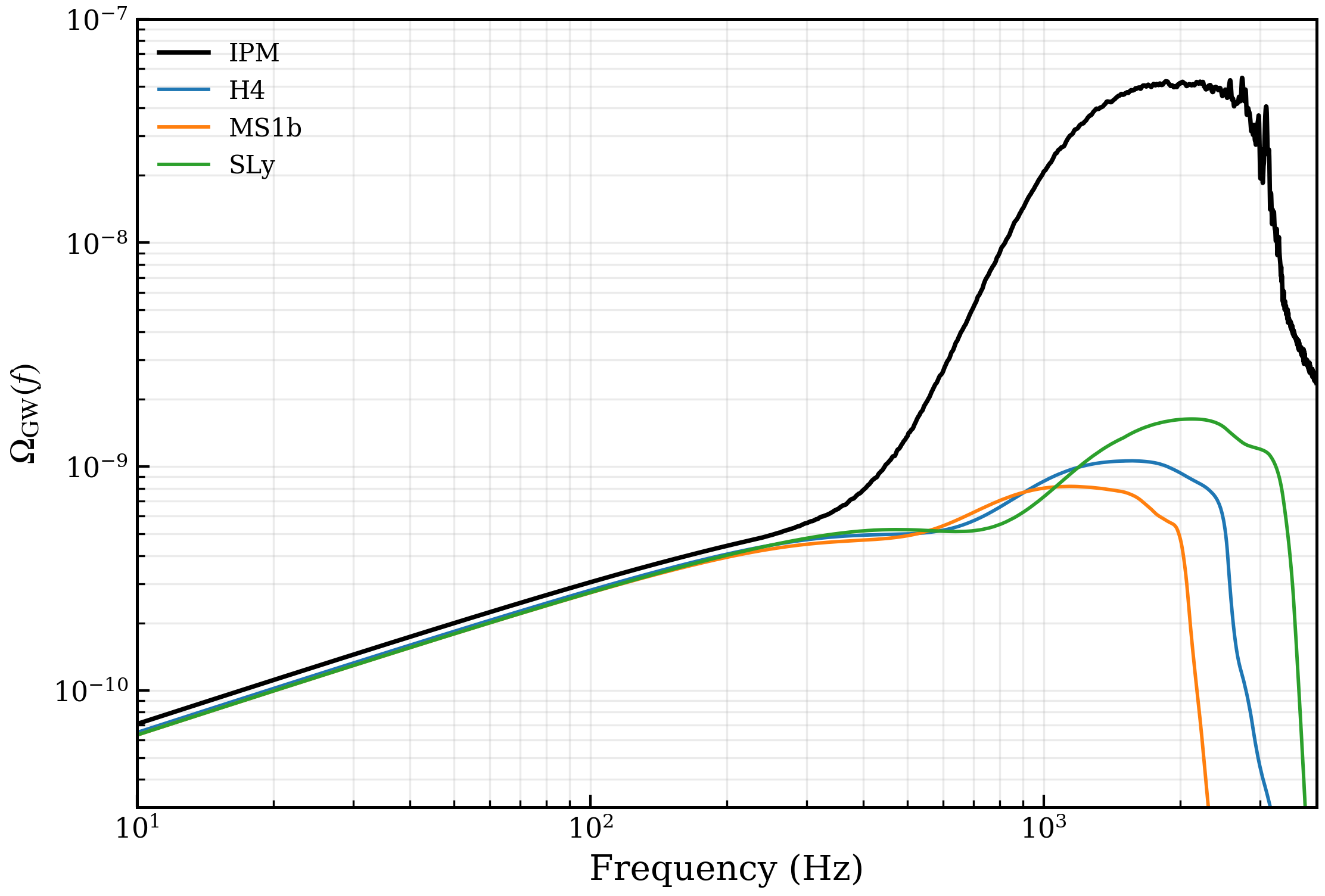}
    \caption{\justifying Comparison of the SGWB spectra predicted by the baseline IPM model and three representative equal-mass, nonspinning NR waveforms, labeled according to their equations of state: SLy, H4, and MS1b. All spectra are shown for a local merger-rate density of $R_0=250~\mathrm{Gpc}^{-3}\,\mathrm{yr}^{-1}$.}
    \label{fig:sgwb_nr_compare}
\end{figure}

\begin{table*}[t]
\centering
\caption{\justifying
SGWB cross-correlation SNRs for the NR comparison waveforms. Entries report the broadband SNR $\rho_{\rm SGWB}^{\mathcal{B}}$, with the PM-band SNR $\rho_{\rm SGWB}^{\mathcal{P}}$ shown in parentheses. As in the baseline calculation, the nearly identical PM-band SNRs for the physical networks reflect the dominance of the internal ET--ET baselines at kilohertz frequencies.
}
\label{tab:sgwb_snr_nr}
\begin{tabular}{llcc}
\hline
\textbf{EoS} & \textbf{Network} &
\textbf{$R_0=7.6$} & \textbf{$R_0=250$} \\
\hline
SLy & ET                  & 3.81 ($6.65\times10^{-4}$) & 125 (0.0219) \\
SLy & ET+CE20+CE40        & 4.50 ($6.66\times10^{-4}$) & 148 (0.0219) \\
SLy & ET+CE20+CE40+ZT     & 4.82 ($6.66\times10^{-4}$) & 159 (0.0219) \\
SLy & ZT (ORF$=1$)        & 2.22 ($2.26\times10^{-3}$) & 73 (0.0744) \\
\hline
H4  & ET                  & 3.91 ($6.84\times10^{-4}$) & 129 (0.0225) \\
H4  & ET+CE20+CE40        & 4.62 ($6.85\times10^{-4}$) & 152 (0.0225) \\
H4  & ET+CE20+CE40+ZT     & 4.95 ($6.85\times10^{-4}$) & 163 (0.0225) \\
H4  & ZT (ORF$=1$)        & 2.28 ($1.25\times10^{-3}$) & 75 (0.0412) \\
\hline
MS1b & ET                 & 3.81 ($5.86\times10^{-4}$) & 125 (0.0193) \\
MS1b & ET+CE20+CE40       & 4.50 ($5.86\times10^{-4}$) & 148 (0.0193) \\
MS1b & ET+CE20+CE40+ZT    & 4.82 ($5.86\times10^{-4}$) & 159 (0.0193) \\
MS1b & ZT (ORF$=1$)       & 2.22 ($9.15\times10^{-4}$) & 73 (0.0301) \\
\hline
\end{tabular}
\end{table*}

\begin{table*}[t]

\centering
\caption{\justifying Predicted annual rates for individually detectable events from the NR comparison waveforms, labeled by EoS. Entries report $\dot N_{\mathcal{B}}$, with the PM-band candidate rate} $\dot N_{\mathcal{P}}$ in parentheses, in units of $\mathrm{yr}^{-1}$. 
\label{tab:single_event_counts_nr}

\begin{tabular}{llcc}
\hline
\textbf{EoS} & \textbf{Detector} & \textbf{$R_0=7.6$} & \textbf{$R_0=250$} \\
\hline
SLy & ET   & 1423.66 (0.0979) & 46831 (3.22) \\
SLy & ZT   & 167.79 (0.218)  & 5520 (7.16) \\
SLy & CE20 & 212.67 (0.0320) & 6996 (1.05) \\
SLy & CE40 & 2306.64 (0.0865) & 75876 (2.84) \\
\hline
H4 & ET   & 1458.70 (0.0904) & 47983 (2.97) \\
H4 & ZT   & 172.43 (0.131)  & 5672 (4.31) \\
H4 & CE20 & 219.13 (0.0439) & 7208 (1.44) \\
H4 & CE40 & 2359.44 (0.0803) & 77613 (2.64) \\
\hline
MS1b & ET   & 1421.22 (0.0930) & 46751 (3.06) \\
MS1b & ZT   & 166.97 (0.0650)  & 5493 (2.14) \\
MS1b & CE20 & 212.40 (0.0445) & 6987 (1.46) \\
MS1b & CE40 & 2303.94 (0.0842) & 75787 (2.77) \\
\hline
\end{tabular}

\end{table*}

To assess the dependence of the forecasts on the PM waveform prescription, we repeat both calculations using the representative SLy, H4, and MS1b NR waveforms described in Sec.~\ref{sec:waveform}.

Fig.~\ref{fig:sgwb_nr_compare} compares the resulting SGWB spectra with the baseline IPM prediction. The spectra are broadly similar at lower frequencies, where the inspiral contribution dominates, but differ substantially in the kilohertz region associated with PM emission. For all three NR waveforms considered here, the PM-band spectra lie well below the baseline IPM prediction. The high-frequency SGWB prediction is therefore sensitive to the adopted PM waveform model, and the analytic IPM prescription should be regarded as a phenomenological baseline rather than a precise representation of the PM spectrum.

Table~\ref{tab:sgwb_snr_nr} quantifies the effect on the stochastic channel. At $R_0=7.6~\mathrm{Gpc}^{-3}\,\mathrm{yr}^{-1}$, ET+CE20+CE40 and ET+CE20+CE40+ZT yield broadband SNRs of $4.50$--$4.62$ and $4.82$--$4.95$, respectively. Their PM-band SNRs are much smaller. For ET+CE20+CE40+ZT, $\rho_{\rm SGWB}^{\mathcal{P}}=(5.86$--$6.85)\times10^{-4}$ at the low-rate endpoint and $0.0193$--$0.0225$ at $R_0=250~\mathrm{Gpc}^{-3}\,\mathrm{yr}^{-1}$. The NR waveforms therefore substantially reduce the predicted PM-band stochastic contribution relative to the baseline IPM calculation, while leaving it well below the adopted detection threshold.

The single-event results show stronger waveform dependence. As listed in Table~\ref{tab:single_event_counts_nr}, the broadband-detectable rates remain large, particularly for CE40, whereas the PM-band candidate rates are much smaller. ZT yields the largest $\dot N_{\mathcal{P}}$ for SLy and H4, reaching $0.218$ and $0.131~\mathrm{yr}^{-1}$ at the low-rate endpoint and $7.16$ and $4.31~\mathrm{yr}^{-1}$ at the high-rate endpoint, respectively. For MS1b, ET instead yields the largest PM-band rates, $0.0930~\mathrm{yr}^{-1}$ and $3.06~\mathrm{yr}^{-1}$. The detailed detector ranking is therefore waveform dependent under our definition of $\rho_{\rm ev}^{\mathcal{P}}$, which integrates the complete stitched waveform over $1$--$4~\mathrm{kHz}$ rather than isolating the post-merger component alone. Dedicated post-merger forecasts using several representative EoSs are presented in Ref.~\cite{ZHTPhysRevX}.

To complement the fixed-mass NR examples, we compare the IPM with TEOBResumSPA--NRPMw~\cite{Huez2026Kilohertz,Breschi2022NRPMw} using the subset of the $R_0=250\,\mathrm{Gpc}^{-3}\,\mathrm{yr}^{-1}$ catalogue with source-frame total masses between $2.4$ and $3.4\,M_\odot$, corresponding to the NRPMw calibration range. In this subset, the IPM ZT count decreases from the full-catalogue value of 174 to 141, whereas TEOBResumSPA--NRPMw yields only 2 PM-band candidates under the same detector responses and SNR criteria. This comparison highlights the strong waveform dependence of the absolute single-event PM-band yield. Appendix~\ref{app:model_dependence} examines two additional assumptions entering the analytic PM prescription.

For individual events with measurable PM emission, joint inspiral--merger--post-merger analyses can connect progenitor masses and tidal properties to the remnant spectrum, thereby constraining the neutron-star mass--radius relation and high-density EoS~\cite{BausweinJanka2012,TorresRivas2019Postmerger}. For example, in full-spectrum Bayesian mock analyses for ET, Huez et al.~\cite{Huez2026Kilohertz} showed that a single event with a PM-band SNR near their detection threshold could constrain the maximum neutron-star mass and maximum density to precisions of approximately 6\% and 10\%, respectively, at the 90\% credibility level. This illustrates the potential scientific value of even a small number of informative PM detections. The PM contribution to the SGWB instead probes the population-integrated emission and is sensitive to average remnant properties, including the strength and duration of PM radiation, although its interpretation is coupled to the merger history and source distribution~\cite{Phinney2001,Lehoucq2025Postmerger}. The two channels therefore provide complementary information at the individual-system and population levels.

\section{Conclusions}
\label{sec:conclusion}

We have compared two observational channels for BNS PM emission: the SGWB generated by the BNS population and direct PM-band observations of individually resolved mergers. Both calculations use the same astrophysical population model and baseline IPM waveform prescription, enabling a consistent comparison between the two channels. The forecasts were evaluated for representative configurations involving ET, CE, and ZT.

For the stochastic channel, the baseline all-source broadband BNS SGWB is detectable even at the lower end of the adopted local merger-rate interval for the representative detector networks considered. In contrast, the PM-band SGWB remains below the adopted threshold for a one-year observation. Even at the upper merger-rate limit and for an idealized kilohertz detector benchmark with ORF$=1$, its SNR reaches only $\rho_{\rm SGWB}^{\mathcal{P}}=2.24$, below the adopted threshold of $3$.

For individual events, ET and CE are more effective for broadband BNS detection, whereas ZT benefits substantially from its enhanced kilohertz sensitivity. Across the adopted local merger-rate interval, ZT yields approximately $5$--$174$ PM-band candidates per year in the baseline IPM calculation, roughly an order of magnitude more than the broadband-oriented detectors. Improved kilohertz strain sensitivity therefore provides a direct advantage for detecting individual high-frequency PM signals.

The NR and TEOBResumSPA--NRPMw comparisons, together with the appendix diagnostics, show that quantitative PM predictions, including both the PM-band candidate rates and the PM-band SGWB, are sensitive to the adopted waveform and modeling assumptions. The representative NR cases give lower PM-band candidate rates for ZT, spanning approximately $0.07$--$7~\mathrm{yr}^{-1}$. The baseline IPM event counts should therefore be regarded as optimistic, model-dependent forecasts rather than precision rate predictions. These results suggest that detailed constraints on remnant structure and the high-density EoS may rely on a relatively small subset of nearby mergers with measurable kilohertz emission, making accurate full-spectrum waveform models particularly important.

Beyond the PM-band question considered above, our residual-background calculation also indicates a broader implication for stochastic searches. The residual BNS background may constitute a non-negligible astrophysical foreground for future searches for weak cosmological SGWBs. Quantifying this limitation will require more realistic treatments of foreground subtraction and subtraction residuals, and will be important for assessing the ultimate sensitivity of third-generation detector networks to cosmological SGWBs.

\appendix

\section{Additional diagnostics of the analytic post-merger model}
\label{app:model_dependence}

The baseline results are based on a representative analytic IPM prescription, whereas Sec.~\ref{sec:results} includes a comparison with a small set of representative NR waveforms. Here, we perform two additional diagnostic variations within the analytic framework. The first modifies the chirp-mass--tidal-deformability mapping used to determine the PM parameters, whereas the second suppresses the PM component associated with a hypermassive neutron star (HMNS) when representative prompt-collapse thresholds are exceeded. These tests are not intended to provide a systematic uncertainty estimate or to marginalize over the EoS and possible remnant outcomes. Instead, they assess whether the predicted PM-band SGWB and its detectability are particularly sensitive to these baseline modeling assumptions.

\subsection{Dependence on the chirp-mass--tidal-deformability mapping}
\label{app:lambda_mapping}

The baseline IPM calculation uses a central $\mathcal{M}_c-\Lambda$ relation to map the sampled chirp mass onto the tidal deformability entering the analytic PM prescription. As discussed in Sec.~\ref{sec:waveform}, this relation is adopted as a phenomenological baseline and should not be interpreted as representing a specific microphysical EoS.

To examine the dependence on this mapping, we repeat the SGWB calculation using two representative lower- and upper-$\Lambda$ relations motivated by the allowed $\mathcal{M}_c-\Lambda$ region, while keeping the power-law exponent fixed at its baseline value of $-3$. The relations are written as
\begin{equation}
\Lambda
=
a+b
\left(
\frac{\mathcal{M}_c}{M_\odot}
\right)^{-3},
\label{eq:lambda_variants}
\end{equation}
where $(a,b)=(-35,1150)$ for the baseline, $(-50,500)$ for the lower-$\Lambda$ variant, and $(-20,1800)$ for the upper-$\Lambda$ variant.

\begin{table}[t]
\centering
\caption{
SGWB SNRs for three chirp-mass--tidal-deformability mappings, evaluated for the ET+CE20+CE40+ZT network with $R_0=250~\mathrm{Gpc}^{-3}\,\mathrm{yr}^{-1}$.
}
\label{tab:lambda_sensitivity}
\begin{ruledtabular}
\begin{tabular}{lcc}
Mapping & $\rho_{\rm SGWB}^{\mathcal{B}}$ & $\rho_{\rm SGWB}^{\mathcal{P}}$ \\
\hline
Baseline                  & 178 & 0.697 \\
Lower-$\Lambda$ variant   & 178 & 0.536 \\
Upper-$\Lambda$ variant   & 178 & 0.697 \\
\end{tabular}
\end{ruledtabular}
\end{table}

\begin{figure}[t]
    \centering
    \includegraphics[width=1.0\linewidth]{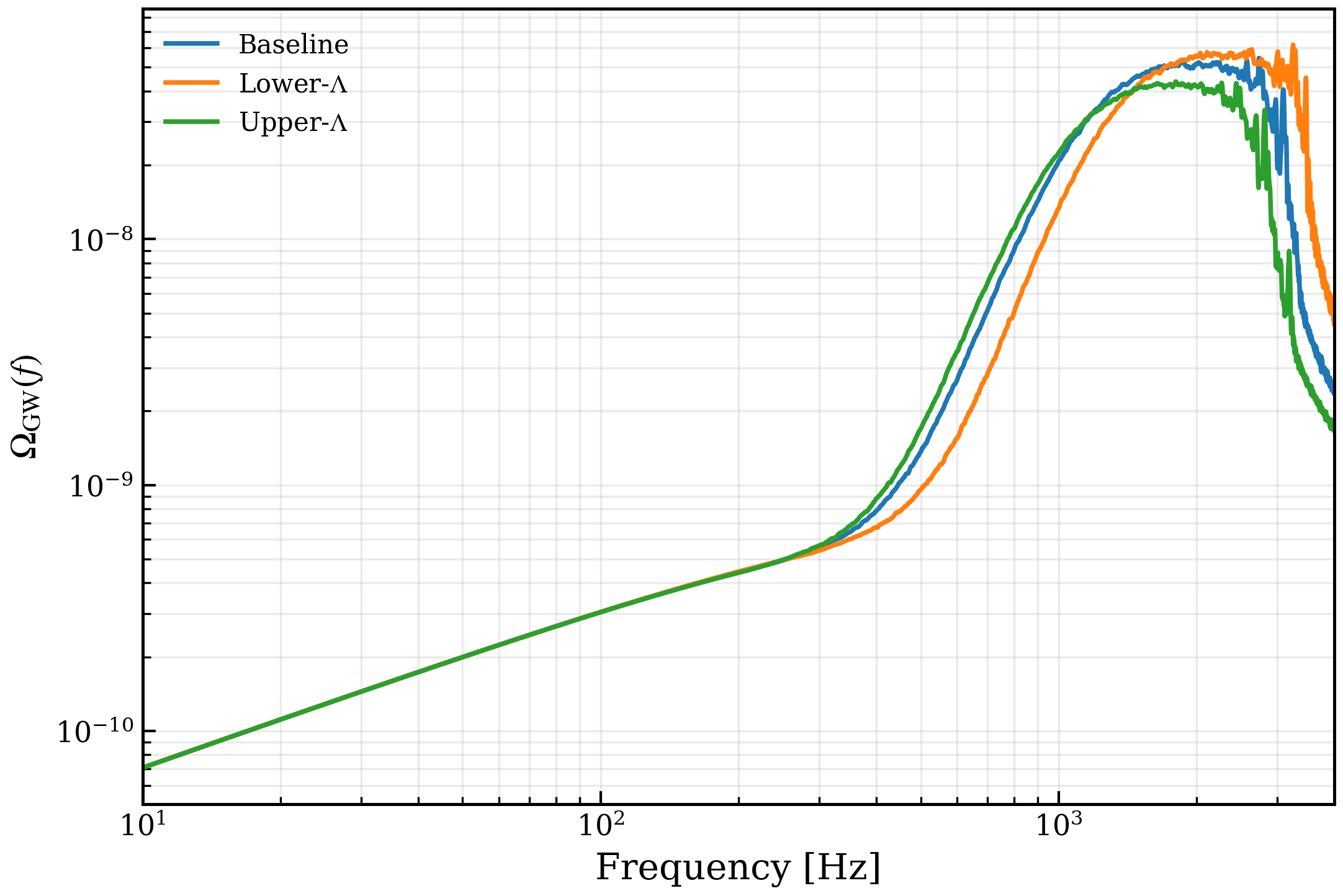}
    \caption{\justifying The SGWB spectra for the baseline and representative lower- and upper-$\Lambda$ mapping variants, evaluated at $R_0=250~\mathrm{Gpc}^{-3}\,\mathrm{yr}^{-1}$.}
    \label{fig:lambda_check}
\end{figure}

Table~\ref{tab:lambda_sensitivity} summarizes the corresponding SNRs. The broadband value remains essentially unchanged because the statistic is dominated by the inspiral--merger contribution. The lower-$\Lambda$ variant reduces $\rho_{\rm SGWB}^{\rm P}$ from $0.697$ to $0.536$. Notably, the baseline and upper-$\Lambda$ mappings yield unrounded PM-band SNRs of $0.696805$ and $0.696847$, respectively, both of which round to $0.697$ at the precision shown in the table. Although their kilohertz spectra differ, as shown in Fig.~\ref{fig:lambda_check}, the changes in their frequency-weighted contributions to the squared SNR nearly cancel after integration.

The tidal mapping therefore affects the normalization and detailed shape of the PM-band contribution, but the resulting stochastic SNR remains well below the adopted $\rho_{\rm SGWB}=3$ threshold for all three mappings considered here.

\subsection{Prompt-collapse diagnostic test}
\label{app:prompt_collapse}

The baseline IPM prescription does not impose an
explicit EoS-dependent prompt-collapse criterion.
Prompt collapse of sufficiently massive BNS systems
to a black hole can strongly suppress or eliminate
the HMNS-like PM emission represented by the
analytic model.
The corresponding threshold is commonly parametrized as
\begin{equation}
M_{\rm thr}=kM_{\rm TOV},
\end{equation}
where $M_{\rm TOV}$ denotes the maximum gravitational mass of a cold, nonrotating neutron star, and the coefficient $k$ depends on the EoS and the binary properties. Numerical-relativity studies generally find thresholds exceeding $M_{\rm TOV}$, with representative values of approximately $k\simeq1.3$--$1.7$~\cite{Bauswein:2013jpa}.

As a diagnostic, rather than an EoS-specific prompt-collapse prescription, we suppress the analytic PM component only for binaries whose source-frame total mass exceeds a representative threshold, $M_{\rm thr}$, while retaining their inspiral--merger contribution. Specifically,
\begin{equation}
\tilde{h}_{\rm PM}(f)
\rightarrow
g(M_{\rm tot}^{\rm src})\,\tilde{h}_{\rm PM}(f),
\label{eq:prompt_suppression}
\end{equation}
where
\begin{equation}
g(M_{\rm tot}^{\rm src})=
\begin{cases}
1, & M_{\rm tot}^{\rm src}\le M_{\rm thr},\\
0, & M_{\rm tot}^{\rm src}>M_{\rm thr}.
\end{cases}
\label{eq:prompt_window}
\end{equation}
We consider $M_{\rm thr}=2.8$, $3.0$, and $3.2\,M_\odot$. These values are not intended to represent precise EoS-dependent collapse thresholds; rather, they are used to bracket the impact of suppressing the analytic PM contribution from the high-mass tail of the adopted population.

\begin{figure}[t]
    \centering
    \includegraphics[width=1.0\linewidth]{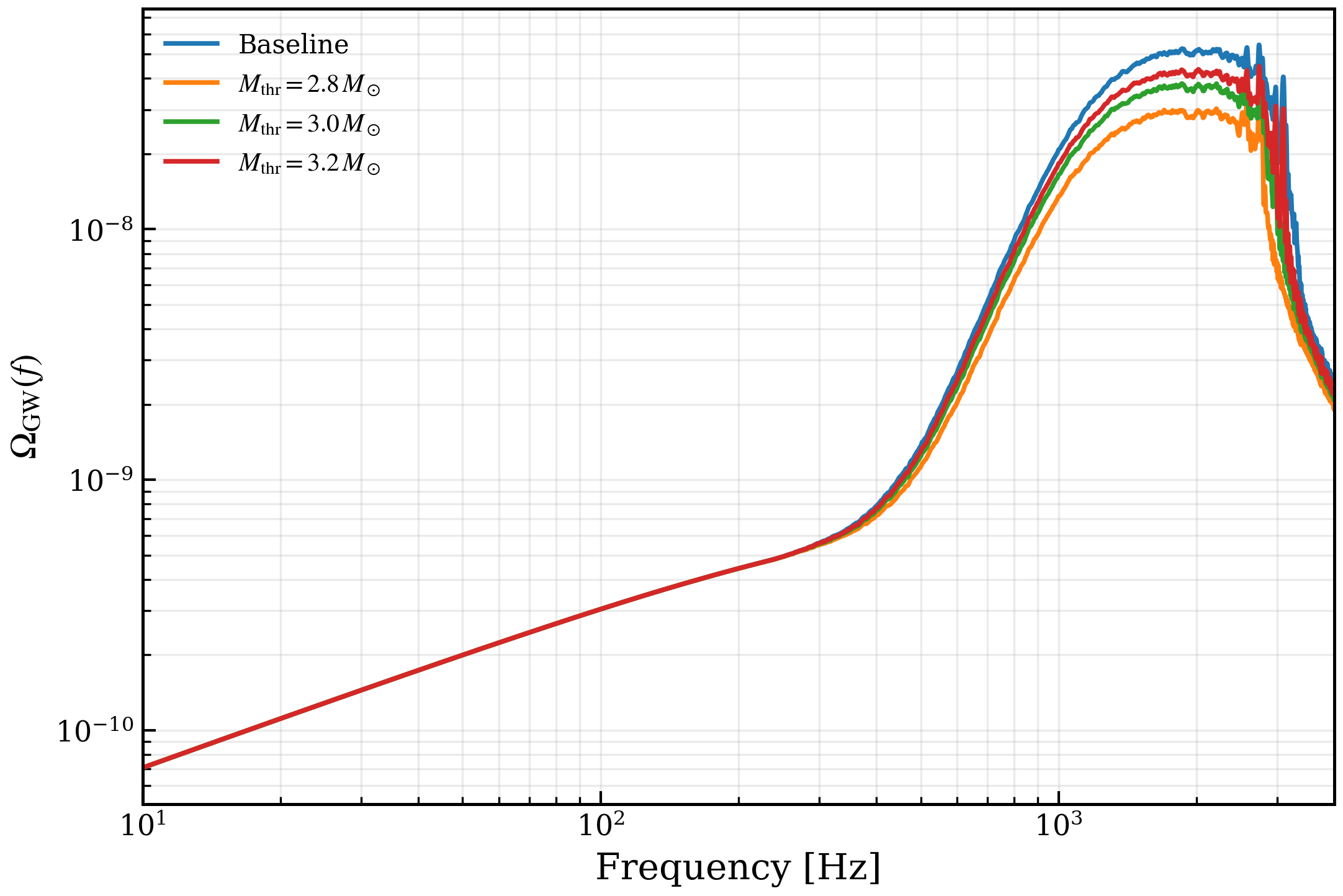}
    \caption{\justifying SGWB spectra for the prompt-collapse diagnostic at $R_0=250~\mathrm{Gpc}^{-3}\,\mathrm{yr}^{-1}$. For binaries with $M_{\rm tot}^{\rm src}>M_{\rm thr}$, only the analytic HMNS-like PM component is suppressed; the inspiral--merger contribution is retained.}
    \label{fig:mass_check}
\end{figure}

\begin{table}[t]
\centering
\caption{
SGWB SNRs for the prompt-collapse diagnostic, evaluated for the ET+CE20+CE40+ZT network at $R_0=250~\mathrm{Gpc}^{-3}\,\mathrm{yr}^{-1}$. For binaries with $M_{\rm tot}^{\rm src}>M_{\rm thr}$, the analytic PM component is suppressed, whereas the inspiral--merger contribution is retained.
}
\label{tab:prompt_suppression}
\begin{ruledtabular}
\begin{tabular}{lcc}
Case & $\rho_{\rm SGWB}^{\mathcal{B}}$ & $\rho_{\rm SGWB}^{\mathcal{P}}$ \\
\hline
Baseline                    & 178 & 0.697 \\
$M_{\rm thr}=2.8\,M_\odot$ & 178 & 0.435 \\
$M_{\rm thr}=3.0\,M_\odot$ & 178 & 0.538 \\
$M_{\rm thr}=3.2\,M_\odot$ & 178 & 0.600 \\
\end{tabular}
\end{ruledtabular}
\end{table}

Table~\ref{tab:prompt_suppression} summarizes the corresponding SNRs, and Fig.~\ref{fig:mass_check} shows the associated spectra. The broadband SNR remains essentially unchanged because the inspiral--merger contribution is retained for all binaries. In contrast, the PM-band SNR decreases monotonically as the assumed prompt-collapse threshold is lowered, from $0.600$ for $M_{\rm thr}=3.2\,M_\odot$ to $0.435$ for $M_{\rm thr}=2.8\,M_\odot$. Relative to the baseline value of $0.697$, these cases correspond to reductions of approximately $14\%$--$38\%$.

Under this diagnostic construction, incorporating prompt-collapse suppression can only reduce the predicted HMNS-like PM contribution. The baseline calculation, which applies the analytic PM prescription to the full simulated population, should therefore be regarded as relatively optimistic for the high-mass tail.

\bibliographystyle{cpc}
\bibliography{BNS}

@article{gwtc4pop,
  archiveprefix = {arXiv},
  author = {Abac, A. G. and Abouelfettouh, I. and Acernese, F. and others},
  collaboration = {LIGO Scientific Collaboration, Virgo Collaboration, and KAGRA Collaboration},
  doi = {10.3847/2041-8213/ae771e},
  eprint = {2508.18083},
  journal = {Astrophys. J. Lett.},
  number = {2},
  pages = {L51},
  primaryclass = {astro-ph.HE},
  title = {GWTC-4.0: Population Properties of Merging Compact Binaries},
  volume = {1005},
  year = {2026}
}

@article{MadauDickinson2014,
  archiveprefix = {arXiv},
  author = {Madau, P. and Dickinson, M.},
  doi = {10.1146/annurev-astro-081811-125615},
  eprint = {1403.0007},
  journal = {Annu. Rev. Astron. Astrophys.},
  pages = {415--486},
  primaryclass = {astro-ph.CO},
  title = {Cosmic Star-Formation History},
  volume = {52},
  year = {2014}
}

@article{Lehoucq2025Postmerger,
  archiveprefix = {arXiv},
  author = {Lehoucq, L{\'e}onard and Dvorkin, Irina and Rezzolla, Luciano},
  doi = {10.1103/4yk2-ltjn},
  eprint = {2503.20877},
  journal = {Phys. Rev. D},
  number = {4},
  pages = {L041306},
  primaryclass = {astro-ph.HE},
  title = {Postmerger emissions: A new and dominant contribution to the gravitational-wave background from binary neutron stars},
  volume = {113},
  year = {2026}
}

@article{Bose2018RadiusPopBNS,
  archiveprefix = {arXiv},
  author = {Bose, Sukanta and Chakravarti, Kabir and Rezzolla, Luciano and Sathyaprakash, B. S. and Takami, Kentaro},
  doi = {10.1103/PhysRevLett.120.031102},
  eprint = {1705.10850},
  journal = {Phys. Rev. Lett.},
  number = {3},
  pages = {031102},
  primaryclass = {gr-qc},
  title = {Neutron-Star Radius from a Population of Binary Neutron Star Mergers},
  volume = {120},
  year = {2018}
}

@article{Altiparmak2022MchirpLambda,
  archiveprefix = {arXiv},
  author = {Altiparmak, Sinan and Ecker, Christian and Rezzolla, Luciano},
  doi = {10.3847/2041-8213/ac9b2a},
  eprint = {2203.14974},
  journal = {Astrophys. J. Lett.},
  number = {2},
  pages = {L34},
  primaryclass = {astro-ph.HE},
  title = {On the Sound Speed in Neutron Stars},
  volume = {939},
  year = {2022}
}

@article{You2025BirthMass,
  archiveprefix = {arXiv},
  author = {You, Zhi-Qiang and Zhu, Xingjiang and Liu, Xiaojin and M{\"u}ller, Bernhard and Heger, Alexander and Stevenson, Simon and Thrane, Eric and Chen, Zu-Cheng and Sun, Ling and Lasky, Paul and Galloway, Duncan K. and Bailes, Matthew and Hobbs, George and Manchester, Richard N. and Gao, He and Zhu, Zong-Hong},
  doi = {10.1038/s41550-025-02487-w},
  eprint = {2412.05524},
  journal = {Nat. Astron.},
  number = {4},
  pages = {552--563},
  primaryclass = {astro-ph.HE},
  title = {Determination of the birth-mass function of neutron stars from observations},
  volume = {9},
  year = {2025}
}

@article{AllenRomano1999,
  archiveprefix = {arXiv},
  author = {Allen, Bruce and Romano, Joseph D.},
  doi = {10.1103/PhysRevD.59.102001},
  eprint = {gr-qc/9710117},
  journal = {Phys. Rev. D},
  number = {10},
  pages = {102001},
  title = {Detecting a stochastic background of gravitational radiation: Signal processing strategies and sensitivities},
  volume = {59},
  year = {1999}
}

@article{RomanoCornish2017,
  author = {Romano, Joseph D. and Cornish, Neil J.},
  doi = {10.1007/s41114-017-0004-1},
  journal = {Living Rev. Relativ.},
  number = {1},
  pages = {2},
  title = {{Detection methods for stochastic gravitational-wave backgrounds: a unified treatment}},
  volume = {20},
  year = {2017}
}

@article{Phinney2001,
  archiveprefix = {arXiv},
  author = {Phinney, E. S.},
  eprint = {astro-ph/0108028},
  title = {A practical theorem on gravitational wave backgrounds},
  year = {2001}
}

@article{Flanagan1993,
  author = {Flanagan, {\'E}anna {\'E}.},
  doi = {10.1103/PhysRevD.48.2389},
  journal = {Phys. Rev. D},
  pages = {2389--2407},
  title = {Sensitivity of the laser interferometer gravitational wave observatory to a stochastic background, and its dependence on the detector orientations},
  volume = {48},
  year = {1993}
}

@article{ZHTPhysRevX,
  archiveprefix = {arXiv},
  author = {Zhang, Teng and Yang, Huan and Martynov, Denis and Schmidt, Patricia and Miao, Haixing},
  doi = {10.1103/PhysRevX.13.021019},
  eprint = {2212.12144},
  journal = {Phys. Rev. X},
  number = {2},
  pages = {021019},
  primaryclass = {gr-qc},
  title = {{Gravitational-Wave Detector for Postmerger Neutron Stars: Beyond the Quantum Loss Limit of the Fabry-Perot-Michelson Interferometer}},
  volume = {13},
  year = {2023}
}

@article{CoReDietrich2018,
  archiveprefix = {arXiv},
  author = {Dietrich, Tim and Radice, David and Bernuzzi, Sebastiano and Zappa, Francesco and Perego, Albino and Br{\"u}gmann, Bernd and Vivekanandji Chaurasia, Swami and Dudi, Reetika and Tichy, Wolfgang and Ujevic, Maximiliano},
  doi = {10.1088/1361-6382/aaebc0},
  eprint = {1806.01625},
  journal = {Class. Quantum Grav.},
  number = {24},
  pages = {24LT01},
  primaryclass = {gr-qc},
  title = {CoRe database of binary neutron star merger waveforms},
  volume = {35},
  year = {2018}
}

@article{Abbott2017GW170817,
  archiveprefix = {arXiv},
  author = {Abbott, B. P. and Abbott, R. and Abbott, T. D. and others},
  collaboration = {LIGO Scientific Collaboration and Virgo Collaboration},
  doi = {10.1103/PhysRevLett.119.161101},
  eprint = {1710.05832},
  journal = {Phys. Rev. Lett.},
  number = {16},
  pages = {161101},
  primaryclass = {gr-qc},
  title = {GW170817: Observation of Gravitational Waves from a Binary Neutron Star Inspiral},
  volume = {119},
  year = {2017}
}

@article{Abbott2017MMA,
  archiveprefix = {arXiv},
  author = {Abbott, B. P. and Abbott, R. and Abbott, T. D. and others},
  doi = {10.3847/2041-8213/aa91c9},
  eprint = {1710.05833},
  journal = {Astrophys. J. Lett.},
  number = {2},
  pages = {L12},
  primaryclass = {astro-ph.HE},
  title = {Multi-messenger Observations of a Binary Neutron Star Merger},
  volume = {848},
  year = {2017}
}

@article{Bauswein2012EoS,
  archiveprefix = {arXiv},
  author = {Bauswein, Andreas and Janka, Hans-Thomas and Hebeler, Kai and Schwenk, Achim},
  doi = {10.1103/PhysRevD.86.063001},
  eprint = {1204.1888},
  journal = {Phys. Rev. D},
  number = {6},
  pages = {063001},
  primaryclass = {astro-ph.SR},
  title = {Equation-of-state dependence of the gravitational-wave signal from the ring-down phase of neutron-star mergers},
  volume = {86},
  year = {2012}
}

@article{Stergioulas2011,
  archiveprefix = {arXiv},
  author = {Stergioulas, Nikolaos and Bauswein, Andreas and Zagkouris, Kimon and Janka, Hans-Thomas},
  doi = {10.1111/j.1365-2966.2011.19488.x},
  eprint = {1105.0368},
  journal = {Mon. Not. R. Astron. Soc.},
  pages = {427--436},
  primaryclass = {astro-ph.SR},
  title = {Gravitational waves and nonaxisymmetric oscillation modes in mergers of compact object binaries},
  volume = {418},
  year = {2011}
}

@article{Bernuzzi2015Spectrum,
  archiveprefix = {arXiv},
  author = {Bernuzzi, Sebastiano and Dietrich, Tim and Nagar, Alessandro},
  doi = {10.1103/PhysRevLett.115.091101},
  eprint = {1504.01764},
  journal = {Phys. Rev. Lett.},
  number = {9},
  pages = {091101},
  primaryclass = {gr-qc},
  title = {Modeling the complete gravitational wave spectrum of neutron star mergers},
  volume = {115},
  year = {2015}
}

@article{Clark2016PostMerger,
  archiveprefix = {arXiv},
  author = {Clark, James Alexander and Bauswein, Andreas and Stergioulas, Nikolaos and Shoemaker, Deirdre},
  doi = {10.1088/0264-9381/33/8/085003},
  eprint = {1509.08522},
  journal = {Class. Quantum Grav.},
  number = {8},
  pages = {085003},
  primaryclass = {astro-ph.HE},
  title = {Observing Gravitational Waves from the Post-merger Phase of Binary Neutron Star Coalescence},
  volume = {33},
  year = {2016}
}

@article{Chatziioannou2017PostMerger,
  archiveprefix = {arXiv},
  author = {Chatziioannou, Katerina and Clark, James Alexander and Bauswein, Andreas and Millhouse, Margaret and Littenberg, Tyson B. and Cornish, Neil},
  doi = {10.1103/PhysRevD.96.124035},
  eprint = {1711.00040},
  journal = {Phys. Rev. D},
  number = {12},
  pages = {124035},
  primaryclass = {gr-qc},
  title = {Inferring the post-merger gravitational wave emission from binary neutron star coalescences},
  volume = {96},
  year = {2017}
}

@article{Punturo2010,
  author = {Punturo, Michele and Abernathy, Matthew and Acernese, Fausto and others},
  doi = {10.1088/0264-9381/27/19/194002},
  journal = {Class. Quantum Grav.},
  number = {19},
  pages = {194002},
  title = {The Einstein Telescope: a third-generation gravitational wave observatory},
  volume = {27},
  year = {2010}
}

@article{Maggiore2020ET,
  archiveprefix = {arXiv},
  author = {Maggiore, Michele and Van Den Broeck, Chris and Bartolo, Nicola and others},
  doi = {10.1088/1475-7516/2020/03/050},
  eprint = {1912.02622},
  journal = {J. Cosmol. Astropart. Phys.},
  pages = {050},
  primaryclass = {astro-ph.CO},
  title = {Science Case for the Einstein Telescope},
  volume = {03},
  year = {2020}
}

@article{Reitze2019CE,
  archiveprefix = {arXiv},
  author = {Reitze, David and Adhikari, Rana X. and Ballmer, Stefan and others},
  eprint = {1907.04833},
  journal = {Bull. Am. Astron. Soc.},
  number = {7},
  pages = {035},
  primaryclass = {astro-ph.IM},
  title = {Cosmic Explorer: The U.S. Contribution to Gravitational-Wave Astronomy beyond LIGO},
  volume = {51},
  year = {2019}
}

@article{Evans2021CE,
  archiveprefix = {arXiv},
  author = {Evans, Matthew and Adhikari, Rana X. and Afle, Chaitanya and others},
  eprint = {2109.09882},
  primaryclass = {astro-ph.IM},
  title = {A Horizon Study for Cosmic Explorer: Science, Observatories, and Community},
  year = {2021}
}

@article{Rosado2011,
  archiveprefix = {arXiv},
  author = {Rosado, Pablo A.},
  doi = {10.1103/PhysRevD.84.084004},
  eprint = {1106.5795},
  journal = {Phys. Rev. D},
  number = {8},
  pages = {084004},
  primaryclass = {gr-qc},
  title = {Gravitational wave background from binary systems},
  volume = {84},
  year = {2011}
}

@article{Zhu2011,
  archiveprefix = {arXiv},
  author = {Zhu, Xing-Jiang and Howell, Eric and Regimbau, Tania and Blair, David and Zhu, Zong-Hong},
  doi = {10.1088/0004-637X/739/2/86},
  eprint = {1104.3565},
  journal = {Astrophys. J.},
  number = {2},
  pages = {86},
  primaryclass = {astro-ph.CO},
  title = {Stochastic Gravitational Wave Background from Coalescing Binary Black Holes},
  volume = {739},
  year = {2011}
}

@article{Planck:2018vyg,
  author = {Aghanim, N. and Akrami, Y. and Ashdown, M. and others},
  collaboration = {Planck Collaboration},
  doi = {10.1051/0004-6361/201833910},
  journal = {Astron. Astrophys.},
  pages = {A6},
  title = {Planck 2018 results. VI. Cosmological parameters},
  volume = {641},
  year = {2020}
}

@article{Gupta2024,
  archiveprefix = {arXiv},
  author = {Gupta, Ish and Afle, Chaitanya and Arun, K. G. and others},
  doi = {10.1088/1361-6382/ad7b99},
  eprint = {2307.10421},
  journal = {Class. Quantum Grav.},
  number = {24},
  pages = {245001},
  primaryclass = {gr-qc},
  reportnumber = {CE Document No. P2300019},
  title = {Characterizing Gravitational Wave Detector Networks: From A$^\sharp$ to Cosmic Explorer},
  volume = {41},
  year = {2024}
}

@article{Branchesi:2023mws,
  archiveprefix = {arXiv},
  author = {Branchesi, Marica and Maggiore, Michele and Alonso, David and others},
  doi = {10.1088/1475-7516/2023/07/068},
  eprint = {2303.15923},
  journal = {J. Cosmol. Astropart. Phys.},
  pages = {068},
  primaryclass = {gr-qc},
  reportnumber = {ET-0084A-23},
  title = {Science with the Einstein Telescope: a comparison of different designs},
  volume = {07},
  year = {2023}
}

@article{Flanagan1997snr,
  archiveprefix = {arXiv},
  author = {Flanagan, {\'E}anna {\'E}. and Hughes, Scott A.},
  doi = {10.1103/PhysRevD.57.4535},
  eprint = {gr-qc/9701039},
  journal = {Phys. Rev. D},
  pages = {4535--4565},
  reportnumber = {GRP-456},
  title = {{Measuring gravitational waves from binary black hole coalescences: 1. Signal-to-noise for inspiral, merger, and ringdown}},
  volume = {57},
  year = {1998}
}

@article{Li2019located_zht,
  archiveprefix = {arXiv},
  author = {Li, Yufeng and Fan, Xilong and Gou, Lijun},
  doi = {10.3847/1538-4357/ab4e18},
  eprint = {1910.08310},
  journal = {Astrophys. J.},
  number = {1},
  pages = {28},
  primaryclass = {astro-ph.CO},
  title = {{Constraining the stochastic gravitational wave from string cosmology with current and future high frequency detectors}},
  volume = {887},
  year = {2019}
}

@article{GW150914,
  archiveprefix = {arXiv},
  author = {Abbott, B. P. and Abbott, R. and Abbott, T. D. and others},
  collaboration = {LIGO Scientific Collaboration and Virgo Collaboration},
  doi = {10.1103/PhysRevLett.116.061102},
  eprint = {1602.03837},
  journal = {Phys. Rev. Lett.},
  number = {6},
  pages = {061102},
  primaryclass = {gr-qc},
  reportnumber = {LIGO-P150914},
  title = {Observation of Gravitational Waves from a Binary Black Hole Merger},
  volume = {116},
  year = {2016}
}

@article{Abbott:2017xzg,
  archiveprefix = {arXiv},
  author = {Abbott, B. P. and Abbott, R. and Abbott, T. D. and others},
  collaboration = {LIGO Scientific Collaboration and Virgo Collaboration},
  doi = {10.3847/2041-8213/aa9a35},
  eprint = {1710.09320},
  journal = {Astrophys. J. Lett.},
  number = {1},
  pages = {L16},
  primaryclass = {astro-ph.HE},
  reportnumber = {LIGO-P1700318},
  title = {Search for Post-merger Gravitational Waves from the Remnant of the Binary Neutron Star Merger GW170817},
  volume = {851},
  year = {2017}
}

@article{Miao:2017iqq,
  archiveprefix = {arXiv},
  author = {Miao, Haixing and Yang, Huan and Martynov, Denis},
  doi = {10.1103/PhysRevD.98.044044},
  eprint = {1712.07345},
  journal = {Phys. Rev. D},
  number = {4},
  pages = {044044},
  primaryclass = {astro-ph.IM},
  title = {{Towards the Design of Gravitational-Wave Detectors for Probing Neutron-Star Physics}},
  volume = {98},
  year = {2018}
}

@article{LIGO170817pm_search,
  archiveprefix = {arXiv},
  author = {Abbott, B. P. and Abbott, R. and Abbott, T. D. and others},
  collaboration = {LIGO Scientific Collaboration and Virgo Collaboration},
  doi = {10.3847/1538-4357/ab0f3d},
  eprint = {1810.02581},
  journal = {Astrophys. J.},
  number = {2},
  pages = {160},
  primaryclass = {gr-qc},
  reportnumber = {LIGO-P1800195},
  title = {Search for gravitational waves from a long-lived remnant of the binary neutron star merger GW170817},
  volume = {875},
  year = {2019}
}

@article{Ackley2020NEMO,
  archiveprefix = {arXiv},
  author = {Ackley, K. and Adya, V. B. and Agrawal, P. and others},
  doi = {10.1017/pasa.2020.39},
  eprint = {2007.03128},
  journal = {Publ. Astron. Soc. Aust.},
  pages = {e047},
  primaryclass = {astro-ph.HE},
  title = {Neutron Star Extreme Matter Observatory: A kilohertz-band gravitational-wave detector in the global network},
  volume = {37},
  year = {2020}
}

@article{Zhu2013cbc,
  archiveprefix = {arXiv},
  author = {Zhu, Xing-Jiang and Howell, Eric J. and Blair, David G. and Zhu, Zong-Hong},
  doi = {10.1093/mnras/stt207},
  eprint = {1209.0595},
  journal = {Mon. Not. R. Astron. Soc.},
  number = {1},
  pages = {882--899},
  primaryclass = {gr-qc},
  reportnumber = {LIGO-P1200105-V2},
  title = {{On the gravitational wave background from compact binary coalescences in the band of ground-based interferometers}},
  volume = {431},
  year = {2013}
}

@article{Dominik:2012pd,
  archiveprefix = {arXiv},
  author = {Dominik, Michal and Belczynski, Krzysztof and Fryer, Christopher and Holz, Daniel and Berti, Emanuele and Bulik, Tomasz and Mandel, Ilya and O'Shaughnessy, Richard},
  doi = {10.1088/0004-637X/759/1/52},
  eprint = {1202.4901},
  journal = {Astrophys. J.},
  number = {1},
  pages = {52},
  primaryclass = {astro-ph.HE},
  title = {{Double Compact Objects I: The Significance of the Common Envelope on Merger Rates}},
  volume = {759},
  year = {2012}
}

@article{Khan2016IMRPhenomD,
  author = {Khan, Sebastian and Husa, Sascha and Hannam, Mark and Ohme, Frank and P{\"u}rrer, Michael and Jim{\'e}nez Forteza, Xisco and Boh{\'e}, Alejandro},
  doi = {10.1103/PhysRevD.93.044007},
  journal = {Phys. Rev. D},
  number = {4},
  pages = {044007},
  title = {Frequency-domain gravitational waves from nonprecessing black-hole binaries. II. A phenomenological model for the advanced detector era},
  volume = {93},
  year = {2016}
}

@article{Dietrich2019NRTidalv2,
  author = {Dietrich, Tim and Samajdar, Anuradha and Khan, Sebastian and Johnson-McDaniel, Nathan K. and Dudi, Reetika and Tichy, Wolfgang},
  doi = {10.1103/PhysRevD.100.044003},
  journal = {Phys. Rev. D},
  number = {4},
  pages = {044003},
  title = {Improving the NRTidal model for binary neutron star systems},
  volume = {100},
  year = {2019}
}

@software{PyCBCSoftware,
  author = {Nitz, Alexander H. and Harry, Ian W. and Brown, Duncan A. and others},
  doi = {10.5281/zenodo.10473621},
  publisher = {Zenodo},
  title = {gwastro/pycbc: v2.3.3 release of PyCBC},
  year = {2024}
}

@article{Bauswein:2013jpa,
  author = {Bauswein, Andreas and Baumgarte, Thomas W. and Janka, Hans-Thomas},
  title = {Prompt merger collapse and the maximum mass of neutron stars},
  journal = {Phys. Rev. Lett.},
  volume = {111},
  number = {13},
  pages = {131101},
  year = {2013},
  doi = {10.1103/PhysRevLett.111.131101},
  eprint = {1307.5191},
  archiveprefix = {arXiv},
  primaryclass = {astro-ph.SR}
}

@article{Thrane:2013oya,
  author = {Thrane, Eric and Romano, Joseph D.},
  title = {Sensitivity curves for searches for gravitational-wave backgrounds},
  eprint = {1310.5300},
  archiveprefix = {arXiv},
  primaryclass = {astro-ph.IM},
  doi = {10.1103/PhysRevD.88.124032},
  journal = {Phys. Rev. D},
  volume = {88},
  number = {12},
  pages = {124032},
  year = {2013}
}

@article{Essick2017FiniteArm,
  author = {Essick, Reed and Vitale, Salvatore and Evans, Matthew},
  title = {Frequency-dependent responses in third generation gravitational-wave detectors},
  journal = {Phys. Rev. D},
  volume = {96},
  number = {8},
  pages = {084004},
  year = {2017},
  doi = {10.1103/PhysRevD.96.084004},
  eprint = {1708.06843},
  archiveprefix = {arXiv},
  primaryclass = {gr-qc}
}

@article{Zhong2023Cosmological,
  author        = {Zhong, Haowen and Ormiston, Rich and Mandic, Vuk},
  title         = {Detecting cosmological gravitational wave background after removal of compact binary coalescences in future gravitational wave detectors},
  journal       = {Phys. Rev. D},
  volume        = {107},
  number        = {6},
  pages         = {064048},
  year          = {2023},
  doi           = {10.1103/PhysRevD.107.064048},
  eprint        = {2209.11877},
  archivePrefix = {arXiv},
  primaryClass  = {gr-qc}
}

@article{Sasli:2023wpa,
    author = "Sasli, Argyro and Karnesis, Nikolaos and Stergioulas, Nikolaos",
    title = "{Exploring the potential for detecting rotational instabilities in binary neutron star merger remnants with gravitational wave detectors}",
    eprint = "2311.10626",
    archivePrefix = "arXiv",
    primaryClass = "gr-qc",
    reportNumber = "109",
    doi = "10.1103/PhysRevD.109.043045",
    journal = "Phys. Rev. D",
    volume = "109",
    number = "4",
    pages = "043045",
    year = "2024"
}

@article{Dietrich2021BNS,
  author        = {Dietrich, Tim and Hinderer, Tanja and Samajdar, Anuradha},
  title         = {Interpreting binary neutron star mergers: describing the binary neutron star dynamics, modelling gravitational waveforms, and analyzing detections},
  journal       = {Gen. Relativ. Gravit.},
  volume        = {53},
  pages         = {27},
  year          = {2021},
  doi           = {10.1007/s10714-020-02751-6},
  eprint        = {2004.02527},
  archivePrefix = {arXiv},
  primaryClass  = {gr-qc}
}

@article{BausweinJanka2012,
  author        = {Bauswein, Andreas and Janka, Hans-Thomas},
  title         = {Measuring Neutron-Star Properties via Gravitational Waves from Neutron-Star Mergers},
  journal       = {Phys. Rev. Lett.},
  volume        = {108},
  pages         = {011101},
  year          = {2012},
  doi           = {10.1103/PhysRevLett.108.011101},
  eprint        = {1106.1616},
  archivePrefix = {arXiv},
  primaryClass  = {astro-ph.SR}
}

@article{TorresRivas2019Postmerger,
  author        = {Torres-Rivas, Andoni and Chatziioannou, Katerina and Bauswein, Andreas and Clark, James Alexander},
  title         = {Observing the post-merger signal of GW170817-like events with improved gravitational-wave detectors},
  journal       = {Phys. Rev. D},
  volume        = {99},
  pages         = {044014},
  year          = {2019},
  doi           = {10.1103/PhysRevD.99.044014},
  eprint        = {1811.08931},
  archivePrefix = {arXiv},
  primaryClass  = {gr-qc}
}

@article{Huez2026Kilohertz,
  author        = {Huez, Giulia and Bernuzzi, Sebastiano and Breschi, Matteo and Gamba, Rossella},
  title         = {Kilohertz gravitational waves from binary neutron star mergers: Full spectrum analyses and high-density constraints on neutron star matter},
  journal       = {Phys. Rev. D},
  volume        = {113},
  pages         = {043019},
  year          = {2026},
  doi           = {10.1103/2jxk-1bf1},
  eprint        = {2507.06293},
  archivePrefix = {arXiv},
  primaryClass  = {gr-qc}
}

@article{Breschi2022NRPMw,
    author = "Breschi, Matteo and Bernuzzi, Sebastiano and Chakravarti, Kabir and Camilletti, Alessandro and Prakash, Aviral and Perego, Albino",
    title = "{Kilohertz gravitational waves from binary neutron star mergers: Numerical-relativity informed postmerger model}",
    eprint = "2205.09112",
    archivePrefix = "arXiv",
    primaryClass = "gr-qc",
    doi = "10.1103/PhysRevD.109.064009",
    journal = "Phys. Rev. D",
    volume = "109",
    number = "6",
    pages = "064009",
    year = "2024"
}

\end{document}